\documentclass[pra,twocolumn,showkeywords,longbibliography,10pt]{revtex4-1}
\usepackage{graphicx}
\usepackage{dcolumn}
\usepackage{color}
\usepackage{times}
\usepackage{bm}
\usepackage{amssymb}
\usepackage{amsmath}
\usepackage{epsfig}
\usepackage{epstopdf}
\usepackage{dsfont}
\usepackage{subfigure}
\usepackage{tikz}
\usepackage[colorlinks, citecolor=blue, linkcolor=blue,urlcolor=blue]{hyperref}
\usepackage[mathscr]{euscript}
\usepackage{orcidlink}
\usepackage{comment}
\usepackage{appendix}
\makeatletter
\renewcommand{\maketag@@@}[1]{\hbox{\m@th\normalsize\normalfont#1}}
\makeatother
\begin{document}
	\renewcommand{\thefootnote}{\fnsymbol {footnote}}
	
	\title{\textcolor{black}{Universal photon blockade via two-photon light-matter interaction  at chiral exceptional points}}

	\author{Hai-Tao Dong}
	\affiliation{State Key Laboratory of Opto-Electronic Information Acquisition and Protection Technology, School of Physics, Anhui University, Hefei
		230601,  People's Republic of China}
	
	\author{Meng-Long Song}
	\affiliation{State Key Laboratory of Opto-Electronic Information Acquisition and Protection Technology, School of Physics, Anhui University, Hefei
		230601,  People's Republic of China}
    
    \author{Si-Yu Zhang}
	\affiliation{State Key Laboratory of Opto-Electronic Information Acquisition and Protection Technology, School of Physics, Anhui University, Hefei
		230601,  People's Republic of China}

	\author{Xue-Ke Song} 
	\affiliation{State Key Laboratory of Opto-Electronic Information Acquisition and Protection Technology, School of Physics, Anhui University, Hefei
		230601,  People's Republic of China}
        
	\author{Liu Ye} 
	\affiliation{State Key Laboratory of Opto-Electronic Information Acquisition and Protection Technology, School of Physics, Anhui University, Hefei
		230601,  People's Republic of China}
	
	\author{Dong Wang\orcidlink{0000-0002-0545-6205}} \email{dwang@ahu.edu.cn}
	\affiliation{State Key Laboratory of Opto-Electronic Information Acquisition and Protection Technology, School of Physics, Anhui University, Hefei
		230601,  People's Republic of China}

	\date{\today}
	
	\begin{abstract}
    The photon blockade (PB) effect is a hallmark non-classical phenomenon in quantum optics and finds important applications for building quantum sources, while the control of PB by the non-Hermitian exceptional points remains largely unexplored. In this work, we theoretically investigate universal photon blockade in a microcavity harboring chiral exceptional points (CEPs) for building multiplexing quantum sources with nonreciprocal photon statistics. The results reveal that the presence of the CEPs leads to a stark contrast in the photon statistics of two whispering-gallery modes with opposite propagating directions. That is, one mode exhibits a strong PB effect while the other displays either sub-Poissonian or super-Poissonian distribution. Our findings thus may pave the way for advanced applications of photon blockade, and provide a theoretical foundation for the selective generation of single-photon and two-photon emission.
	\end{abstract}
	
	\maketitle

	\textcolor{black}{\section{Introduction}}
	Photon blockade (PB) is a typical non-classical correlation effect within an optical cavity system \cite{1,2,3,4,5,6,7,8}, where the first photon  blocks the excitation of subsequent photons \cite{9}. PB not only takes an important role in the quantum control of optical fields and photon statistical properties, but also  be applied to the achievement of quantum light sources \cite{10,11,12,13,14,15,16,17,18,19}. Generally, PB can be classified into two types \cite{1,2,20,21,22}: conventional photon blockade (CPB) and unconventional photon blockade (UPB). CPB originates from the strong anharmonic nature of the system's energy spectrum, with energy levels having unequal spacing, causing severe detuning of two-photon transitions and thereby inhibiting multi-photon excitation \cite{1}. UPB arises from quantum interference between different excitation pathways, achieving photon anti-bunching \cite{23,24,25,26,27}. Theoretically, we can  obtain the characteristics of PB through photon antibunching and sub-Poisson photon number statistics \cite{28,29}.

    \begin{figure*}
		
			\centering
			\begin{minipage}{1.0\textwidth}
			\centering
		\subfigure{\includegraphics[height=5cm]{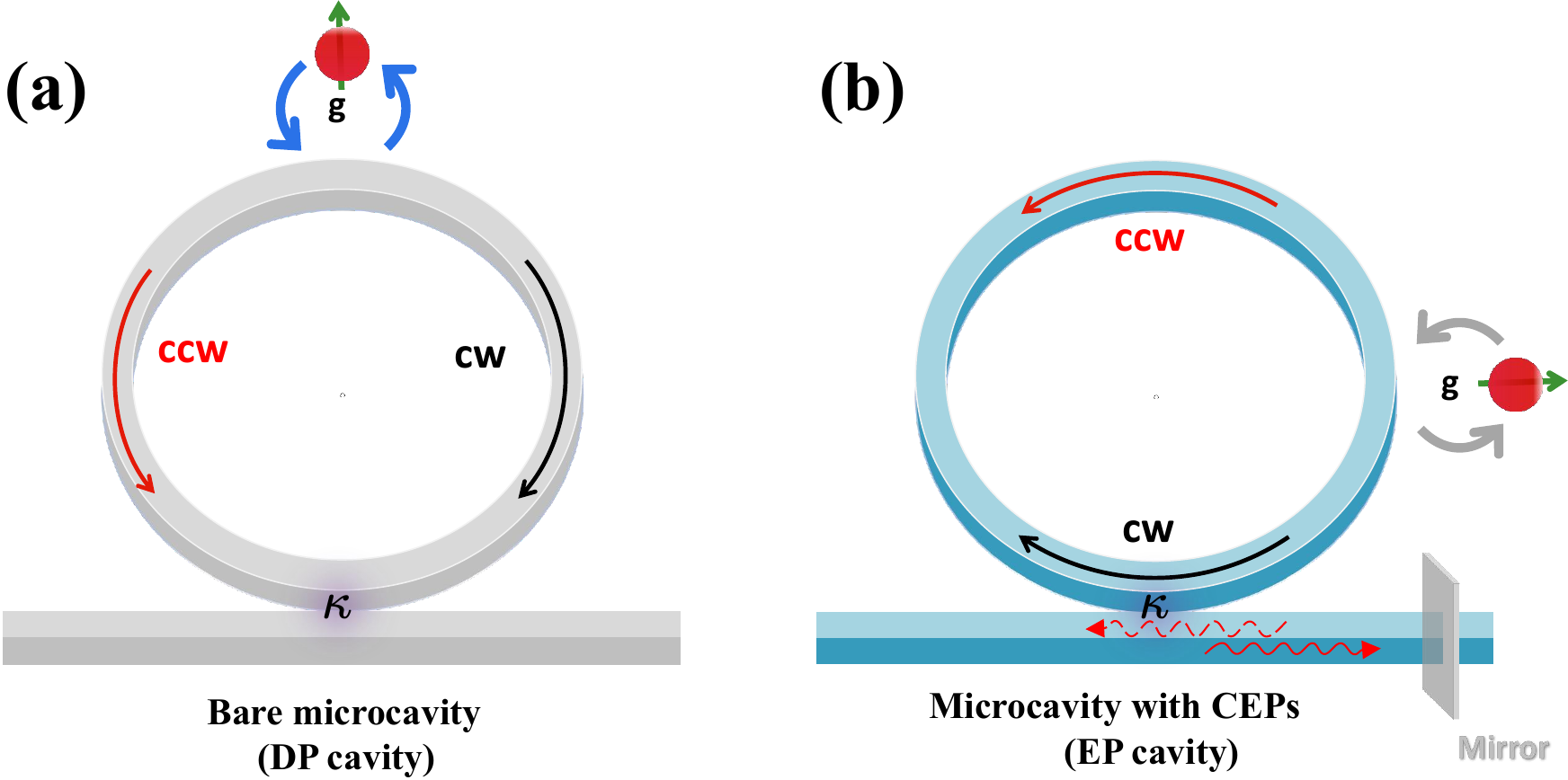}}
		\subfigure{\includegraphics[height=6cm]{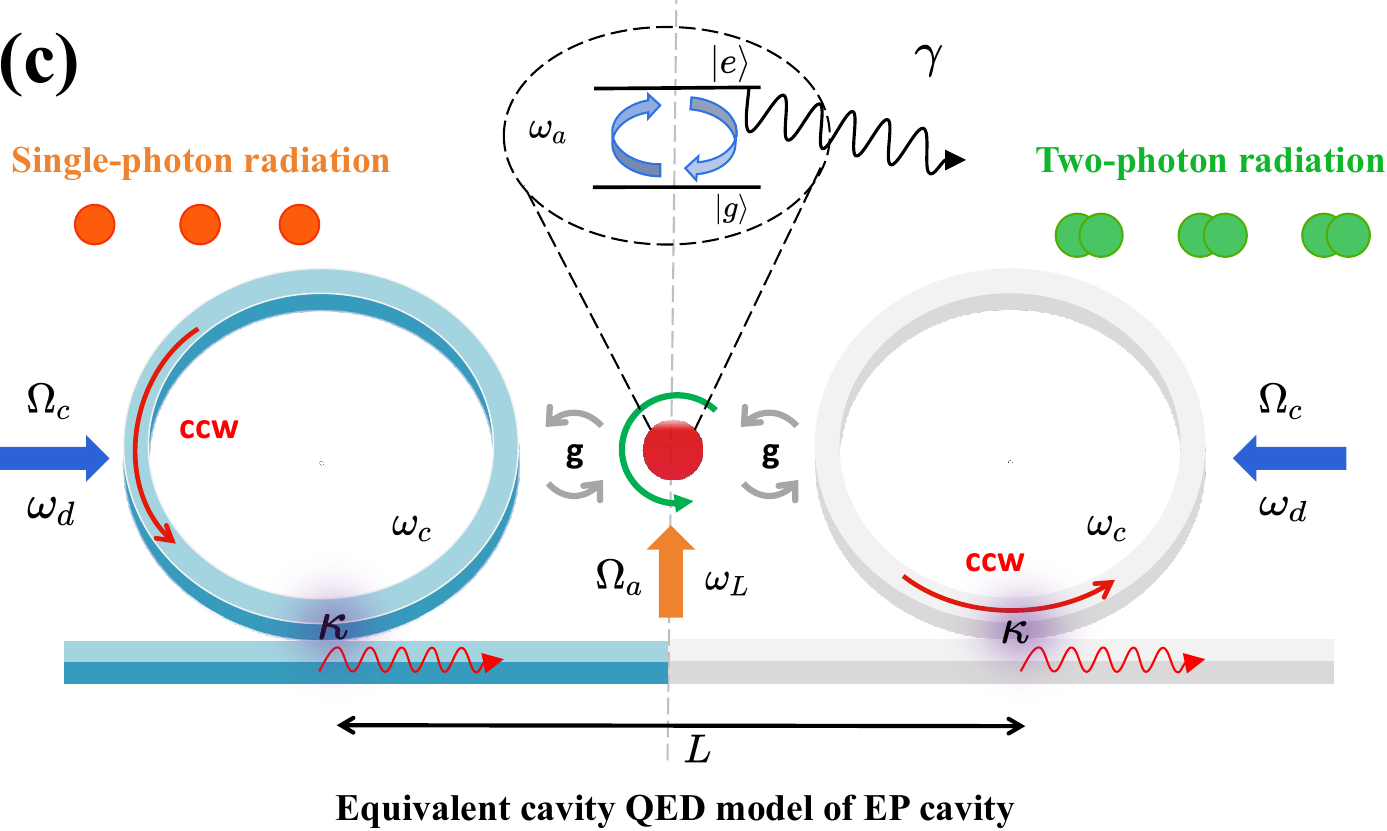}}

		\end{minipage}\hfill
		\caption{Schematics of the cavity QED systems and the equivalent models. (a) A standard resonator supporting degenerate CW and CCW modes. A two-level atom couples to both modes with equal strength $g$. $\kappa$ denotes the cavity-waveguide decay rate. (b)  A mirror on the waveguide introduces unidirectional coupling from the CCW to the CW mode, driving the system to CEPs. (c) The EP system is unfolded via mirror symmetry into a cascaded model with a spatial separation $L$. $\Omega_{c} $ ($\Omega_{a}$) represents the driving strength of the pump field, and $\omega_d$ ($\omega_L$) represents the driving frequency.}
		\label{fig1}
	\end{figure*}
\textcolor{black}{While both mechanisms exhibit these statistical features, their distinct physical origins lead to different practical limitations. As a matter of fact, CPB provides a robust blockade but strictly requires strong nonlinearity, which is experimentally challenging. By contrast, UPB requires much weaker nonlinearity, but is highly parameter-sensitive and operates within a restricted space. To overcome these limitations and utilize their advantages, recent studies have explored the intrinsic connection between CPB and UPB to establish a more integrated framework.} This leads to the concept of universal photon blockade, where photon antibunching can be achieved across the entire range from weak to strong coupling. Within this evolving field,  Zhou {\it  et al.} \cite{30} recently derived a set of unified optimal analytical conditions for achieving universal photon blockade by employing a two-photon Jaynes-Cummings (JC) model \cite{31,32,33}. Under this framework, the equal-time second-order correlation function $g^{(2)}(0)$ can be minimized regardless of whether the system operates in the strong nonlinearity regime, the weak nonlinearity regime, or the intermediate transition region. This universal framework drastically broadens the parameter space for modulating quantum light sources.

 Beyond CPB and UPB, alternative mechanisms such as the quantum Zeno blockade have also been proposed to dynamically restrict a bosonic Hilbert space to finite-dimensional subspaces \cite{34}. Theoretical investigations into PB are well-established.  However, cavity loss poses a major challenge to its efficacy in real-world systems. Interestingly, such losses can be harnessed as a resource for state modulation by exploiting exceptional points (EPs) \cite{35,36,37,38,39,40}, a central concept of non-Hermitian Physics. Traditional EPs are characterized by the coalescence  of eigenvalues and eigenstates, whereas the introduction of unidirectional coupling gives rise to chiral exceptional points (CEPs) \cite{41,42,43,44,45,46}. Optical microcavities serve as ideal non-Hermitian platforms for engineering CEPs due to their inherent dissipation and high-quality factors. In whispering-gallery-mode geometries, the degeneracy of clockwise (CW) and counter-clockwise (CCW) modes provides the essential basis for introducing chirality \cite{47}. \textcolor{black}{ To construct CEPs via the interaction of two modes, several physical strategies have been explored, such as introducing Rayleigh-type nano-scatterers to break the CW-CCW symmetry \cite{48}, employing an external nanotip for local field manipulation \cite{43}, or coupling the resonator to a waveguide with a terminal mirror \cite{44}. }
 
Compared to the fabrication and stability challenges associated with nano-scatterers and external nanotips, the waveguide with a terminal mirror provides superior in-situ tunability and precise phase control. Driven by its distinct advantages for stable on-chip integration, we adopt this latter configuration to construct CEPs and propose the universal photon blockade in a microcavity system \cite{45,46,47,48,49}. \textcolor{black}{To achieve PB within this non-Hermitian system, we introduce a two-photon light-matter interaction mechanism described by a two-photon two-mode JC model. This model inherently possesses stronger nonlinearity, and the resulting multi-pathway quantum interference effectively relaxes conventional strong-coupling requirements. This not only lays the physical foundation for universal photon blockade but also provides an ideal framework for selectively modulating single-photon and two-photon emissions.} Besides, we examine the influence of cavity and atom driven schemes on the photon statistical properties of the system. Importantly, the optimal analytical conditions are derived  for universal photon blockade, which are also verified by numerical simulations. Our findings reveal that under cavity driving, the system exhibits completely different photon statistics for its two modes when the weak driving field is resonant with the cavity frequency: the photon statistics of the CCW mode follow a Poissonian distribution, whereas the  CW mode displays a pronounced PB. Conversely, in the detuned regime, this result is reversed—the CW mode shifts toward Poissonian or even bunching statistics, while the CCW mode exhibits strong PB. In addition, both the CW and CCW modes simultaneously exhibit pronounced photon bunching in the regime of the atom-driven. With these in mind, the current work provides insight into PB in the framework of two-photon light-matter interaction at chiral exceptional points, and will be beneficial for realizing directionally controllable single-photon sources and nonreciprocal quantum optical components.

{\section{ MODEl and Theory}}

We exam a model of the interaction between the quantum emitter (QE) and the microcavity with CEPs. A bare microcavity, as depicted in Fig. \ref{fig1}(a), supports two degenerate traveling-wave modes: the CW and CCW modes. In the absence of direct coupling between them, the cavity operates at a diabolic point (DP), where the eigenvalues are degenerate while the corresponding eigenstates remain strictly orthogonal. The microcavity with the CEPs is depicted in Fig. \ref{fig1}(b). For visualization of the model, we construct an equivalent model of Fig. \ref{fig1}(b) utilizing the  mirror symmetry, as depicted in Fig. \ref{fig1}(c). The CW mode in the original system is flipped into a mirrored CCW mode, thereby mapping the originally complex closed-loop feedback path onto an intuitive one-dimensional cascaded system. In this configuration, photons in the waveguide exhibit strict chiral characteristics, permitting unidirectional transport exclusively from the left cavity ($c_L$) to the right cavity ($c_R$).

Basically, the two-photon interaction between the QE and photons can be described by a two-photon two-mode JC model \cite{50,51,52}
\begin{equation}
    	H_{\rm JC} = H_0 + H_{I},
		\label{Eq.1}
\end{equation}
with the free Hamiltonian ($\hbar=1$)
\begin{equation}
		H_0 = \omega_c c_{L}^{\dagger}c_{L}+\omega_c c_{R}^{\dagger}c_{R}+\omega_{a}\sigma_{+}\sigma_{-}
		\label{Eq.2}
\end{equation}
and the interaction Hamiltonian 
\begin{equation}
		H_{I} = g(c^{\dagger2}_{L}\sigma_{-}+
        c^{\dagger2}_{R}\sigma_{-}+h.c.).
		\label{Eq.3}
\end{equation}
Where $c_{L} $($c_R$) is the bosonic annihilation operator of the left (right) cavity with a resonance frequency $\omega_c$, where $\sigma_{+}=| e \rangle \langle g | $ and $\sigma_{-}= |g \rangle \langle e|$ represent the ladder operators of the atom with a transition frequency $\omega_a$ between the ground state $|g \rangle$ and excited state $|e \rangle$, where $g$ denotes  the coupling strength between the photon and the atom.

To achieve and manipulate PB, we investigate the PB effect in the system by driving either the cavity, $H_{d} = \Omega_c(c_{L}^{\dagger}e^{-i \omega_dt}+c_{L}e^{i\omega_dt})+\Omega_c(c_{R}^{\dagger}e^{-i\omega_dt}+c_{R}e^{i\omega_dt})$, or the atom, $H_{d}^{'}=\Omega_a(\sigma_{+}e^{-i\omega_{L}t}+\sigma_{-}e^{i\omega_{L}t})$, where $\Omega_c$ ($\Omega_a$)  and $\omega_c$ ($\omega_a$) denote the driving strength and driving frequency of the cavity (atom). The total Hamiltonian of the system can be given by $H_{ t}=H_{\rm JC}+H_{d} (H_{d}^{'})$.

In addition, we use the extended cascade quantum master equation to describe the dynamic evolution of the unidirectionally coupled system under the theoretical framework of the equivalent cascade model \cite{46,53,54,55,56}
\begin{align}
		\dot{\rho}=-i[&H_t,\rho]+\kappa \mathcal{L}[c_L]\rho + \kappa \mathcal{L}
        [c_R]\rho + \gamma\mathcal{L}[\sigma_{-}]\rho \nonumber \\ 
        &+\kappa|r|(e^{i\varphi}[c_{L}\rho,c_{R}^{\dagger}]+e^{-i\varphi}[c_{R},\rho c_{L}^{\dagger}]),
		\label{Eq.4}
\end{align}
where $\kappa$ is the dissipative coupling rate between the microcavity and the waveguide. $\gamma$ denotes the spontaneous emissivity (SE) of atoms in homogeneous medium. $\mathcal{L}[\mathcal{O} ]\rho = \mathcal{O}\rho \mathcal{O}^{\dagger }-\{{\mathcal{O}^{\dagger } \mathcal{O},\rho }\}/2 $ is the  Lindblad superoperator for operator $\mathcal{O}$. Besides, $|r|$ characterizes the reflectivity of mirror, and $\varphi = \beta L$ denotes the phase factor, with $\beta$ and $L$ being the propagation constant of waveguide and the distance between two cavities, respectively.

Based on the master equation, the time evolution of the expectation value for an arbitrary system operator $\mathcal{O}$ can be derived using the Heisenberg-Langevin equation
\begin{align}
		\dot{\left \langle \mathcal{O} \right \rangle}=Tr(\mathcal{O}\dot{\rho}) .
		\label{Eq.5}
\end{align}
By substituting the system operators $c_L$, $c_R$, and $\sigma_-$ into Eq. (\ref{Eq.5}),  we obtain the equations of motion for the system operators
\begin{subequations}\label{eq6}
\begin{align}
\langle\dot{c}_L\rangle &= -i \left[ \left(\omega_c - i\frac{\kappa}{2}\right)\langle c_L \rangle + 2g\langle c_L^\dagger \sigma_- \rangle \right], \\
\langle\dot{c}_R\rangle &= -i \left[ \left(\omega_c - i\frac{\kappa}{2}\right)\langle c_R \rangle - i\kappa|r|e^{i\varphi}\langle c_L \rangle + 2g\langle c_R^\dagger \sigma_- \rangle \right], \\
\langle\dot{\sigma}_-\rangle &= -i \left[ \left(\omega_a - i\frac{\gamma}{2}\right)\langle \sigma_- \rangle + g\langle c_L^2 \rangle + g\langle c_R^2 \rangle \right].
\end{align}
\end{subequations}
The above equations of motion are inherently nonlinear and unclosed, accompanied by a unidirectional coupling induced by the cascaded waveguide.

To quantitatively describe the PB effect and characterize the photon statistical properties of the output field from a theoretical perspective, we introduce the correlation function. In quantum optics, the statistical nature of two cavities are typically described by the equal-time second-order correlation function  \cite{57}
\begin{align}
		g_{ L}^{(2)}(0)=\frac{\langle \hat{c}_{L}^{\dagger}(0)\hat{c}_{L}^{\dagger}(0)\hat{c}_{L}(0)\hat{c}_{L}(0)\rangle}{\langle \hat{c}_{L}^{\dagger}(0)\hat{c}_{L}(0)\rangle^{2}},
		\label{Eq.7}
\end{align}

\begin{align}
		g_{ R}^{(2)}(0)=\frac{\langle \hat{c}_{R}^{\dagger}(0)\hat{c}_{R}^{\dagger}(0)\hat{c}_{R}(0)\hat{c}_{R}(0)\rangle}{\langle \hat{c}_{R}^{\dagger}(0)\hat{c}_{R}(0)\rangle^{2}}.
		\label{Eq.8}
\end{align}

 Based on the value of  $g^{(2)}(0)$, the statistical properties of a light field are strictly categorized into three regimes: Poissonian statistics for $g^{(2)}(0) = 1$, super-Poissonian statistics (photon bunching) for $g^{(2)}(0) > 1$, and non-classical sub-Poissonian statistics (photon anti-bunching) for $g^{(2)}(0) < 1$. In particular, the vanishing limit $g^{(2)}(0) \to 0$ serves as the definitive signature and primary criterion for an ideal single‑photon source enabled by PB.
\\
\\

\textcolor{black}{\section{ UNIVERSAL PHOTON BLOCKADE IN THE CAVITY-FIELD-DRIVING CASE}\label{3h}}

In this section, we analytically calculate the equal-time second-order correlation functions for the left and right cavities under cavity driving schemes. We derive the optimal conditions for universal photon blockade and validate our results through numerical simulations using QuTip \cite{58,59}.

Under continuous-wave monochromatic weak driving, the Hamiltonian of the system becomes $H_{ t}=H_{\rm JC}+H_{d}$. For simplification, we set $\omega_
L = 2\omega_d$. In a rotating frame defined by the unitary operator exp[$-i\omega_d (c_L^{\dagger}c_{L}+c_{R}^{\dagger} c_{R})t-i\omega_L\sigma_{+}\sigma_{-}t$], the Hamiltonian of the system becomes
\begin{align}
		H_{\rm sys}=\Delta_{c}c_{L}^{\dagger}c_L+\Delta_{c}c_{R}^{\dagger}c_R+\Delta_a\sigma_{+}\sigma_{-}+g(c_{L}^{\dagger2}\sigma_{-}+h.c.)\nonumber \\
+g(c_{R}^{\dagger2}\sigma_{-}+h.c.)+\Omega_c(c_L^{\dagger}+c_L)
+\Omega_c(c_R^{\dagger}+c_R) ,
		\label{Eq.9}
	\end{align}
where, $\Delta_{c} = \omega_{c}-\omega_d$ ($\Delta_{a} = \omega_{a}-2\omega_d$) is the detuning of the cavity-field (atomic) frequency with respect to the driving frequency. We here set $\omega_a = 2\omega_c$, thus $\Delta_a=2\Delta_c$ is held.

To incorporate the effects of dissipation between the open system and its environment on PB, we employ an effective non-Hermitian Hamiltonian \cite{60,61,62,63}
\begin{align}
		H_{\rm eff}=H_{\rm sys}-\sum_{s=L,R}^{}i\frac{\kappa }{2}c_{s}^{\dagger }c_s  -i\kappa|r|e^{i\varphi}c_{L}c_{R}^{\dagger}-i\frac{\gamma}{2}\sigma_{+}\sigma_{-}.
		\label{Eq.10}
	\end{align}
    
In the weak-driving regime ($\Omega_c\ll \kappa,\gamma $), we can ignore the influence of the external pump fields. Consequently, the system possesses a conserved quantity, namely the symmetry operator \cite{64,65} $C=\sum_{s}^{}c_{s}^{\dagger }c_s+2\sigma _{+}\sigma _{-}$, where $s=L,R$.
The effective Hamiltonian can be diagonalized in subspaces $\{{\left | 00g \right \rangle}\},\{{\left | 10g  \right \rangle,\left | 01g  \right \rangle }\},
\{{\left | 20g \right \rangle,\left | 02g \right \rangle,\left | 11g \right \rangle,\left | 00e \right \rangle}\}$, $\cdots$,
$\{{\left | n0g \right \rangle,\left | 0ng \right \rangle,\cdots}\}$,
where $n = 1,2,3,\dotsc$ is the photon numbers. Under the weak-driving limit, the probability amplitudes of higher-order Fock states are sufficiently small to be neglected. Therefore, we truncate the Hilbert space of photon numbers to $n=2$. We express the state function of the system in Dirac notation, expanding it in terms of the basis vectors of the subspaces
\begin{align}
|\Psi(t)\rangle &= C_{00g}(t)|00g\rangle + C_{10g}(t)|10g\rangle + C_{01g}(t)|01g\rangle \nonumber  \\&+ C_{11g}(t)|11g\rangle +C_{20g}(t)|20g\rangle \nonumber  + C_{02g}(t)|02g\rangle \\&+ C_{00e}(t)|00e\rangle,
\label{Eq.11}
\end{align}
where, $\left | ijk \right \rangle =\left | i \right \rangle \otimes \left | j \right \rangle
\otimes \left | k  \right \rangle $ represents the product state of the atom and
the cavity. Here $\left | i(j) \right \rangle$ ($i,j=0,1,2$) denotes the Fock state  and $ \left | k \right \rangle$ ($k=g,e$) indicates the atomic state. The coefficients $C_{ijk}(t)$ are the probability amplitudes of $\left | ijk \right \rangle$. Based on the Schr$\ddot{\rm o}$dinger equation $i \partial t \left | \Psi (t) \right \rangle = H_{\rm eff}\left | \Psi(t)  \right \rangle $, we derive the following set of equations of motion for the probability amplitudes:
\begin{subequations}
\begin{align}
i{\dot C}_{00g} &= {\Omega _c}{C_{10g}} + {\Omega _c}{C_{01g}},\label{12.a}\\
i{\dot C}_{10g} &= {\Omega _c}{C_{00g}} + {\tilde \Delta }_c{C_{10g}} + {\Omega _c}{C_{11g}} + \sqrt 2 {\Omega _c}{C_{20g}},\label{12.b} \\
i{\dot C}_{01g} &= {\Omega _c}{C_{00g}} - i\kappa |r|{e^{i\varphi }} {C_{10g}} + {\tilde \Delta }_c{C_{01g}} + {\Omega _c}{C_{11g}}\nonumber \\
&+ \sqrt 2 {\Omega _c}{C_{02g}} ,\label{12.c}\\
i{\dot C}_{11g} &= {\Omega _c}{C_{10g}} + {\Omega _c}{C_{01g}} - \sqrt 2i\kappa |r|{e^{i\varphi }}{C_{20g}}\nonumber\\
&+ 2{\tilde \Delta }_c{C_{11g}},\label{12.d}\\
i{\dot C}_{20g} &=\sqrt 2 {\Omega _c}{C_{10g}}+ 2{\tilde \Delta }_c{C_{20g}}+\sqrt 2gC_{00e},\label{13.e}\\
i{\dot C}_{02g} &=\sqrt 2 {\Omega _c}{C_{01g}}-\sqrt 2i\kappa |r|{e^{i\varphi }}{C_{11g}}+ 2{\tilde \Delta }_c{C_{02g}}\nonumber\\
&+\sqrt{2}gC_{00e},\label{13.f}\\
i{\dot C}_{00e} &=\sqrt{2}gC_{20g}+\sqrt{2}gC_{02g}+{\tilde{\Delta}}_aC_{00e}.\label{13.g}
\end{align}
\end{subequations}
Here, we define $\tilde {\Delta }_c=\Delta _{c}-i\kappa /2 $ and  $\tilde {\Delta }_a=\Delta _{a}-i\gamma/2 $. The higher-order excitation probabilities of the system decay drastically with the effective excitation number. Therefore, the higher-order small terms in the equations of motion can be safely neglected, and the vacuum state amplitude can be approximated as $C_{00g} \approx 1$. By employing a perturbation method \cite{66,67}, we then derive the steady-state solutions for the probability amplitudes
\begin{subequations}
\begin{align}		
  C_{10g} &= -\frac{\Omega_c}{\tilde{\Delta}_c}, \label{13.a}\\ 
  C_{01g} &= -\frac{\Omega_c(\tilde{\Delta}_c+i\kappa|r|e^{i\varphi})}{\tilde{\Delta}^{2}_{c}},\label{13.b} \\ 
  C_{20g} &= \frac{\Omega_{c}^{2}(\Upsilon+2\tilde{\Delta}_a\tilde{\Delta}_c^{3})}{\sqrt{2}\tilde{\Delta}_{c}^{2}\varpi},\label{13.c}\\ 
  C_{02g} &= -\frac{\Omega_{c}^{2}(\Upsilon+\Lambda )}{\sqrt{2}\tilde{\Delta}_{c}^{2}\varpi},\label{13.d} 
\end{align}
\end{subequations}
where $\varpi=g^{2}\kappa^{2}|r|^{2}e^{2i\varphi}-\tilde {\Delta }^{2}_{c}(4g^{2}-2\tilde {\Delta }_{c}\tilde {\Delta }_a)$, $\Upsilon =g^{2}\kappa |r|e^{i\varphi }(4i\tilde {\Delta }_c-\kappa |r|e^{i\varphi})$ and $\Lambda =2\tilde {\Delta }_a\tilde {\Delta }_c(\kappa |r|e^{i\varphi }-i\tilde {\Delta }_c)^{2}$. 
It is worth noting that if there is no coupling term ($|r|=0$), then we can obtain $C_{10g}=C_{01g}$ and $C_{20g}=C_{02g}$, and the results are consistent with the form in Ref. \cite{28}.

 Under the weak driving conditions, Eqs. (\ref{Eq.7}) and (\ref{Eq.8}) can be simplified  as
\begin{align}
		g^{(2)}_L(0)=\frac{{\langle c_L^{\dagger2}c_L^{2}\rangle}}{\langle c_L^{\dagger}c_L\rangle^{2}}\approx\frac{2|C_{20g}|^{2}}{|C_{10g}|^{4}},\\
        g^{(2)}_R(0)=\frac{{\langle c_R^{\dagger2}c_R^{2}\rangle}}{\langle c_R^{\dagger}c_R\rangle^{2}}\approx\frac{2|C_{02g}|^{2}}{|C_{01g}|^{4}}.
		\label{Eq.15,16}
	\end{align}
 \\
\\
   \textcolor{black}{\subsection{ Resonant case }}
  
    We first consider the resonant case $\Delta_c=0$. Similar to Ref. \cite{30}, the optimal conditions for the universal photon blockade are obtained from the requirement that the probability amplitude of the two-photon Fock state vanishes. For the right cavity, it satisfies 
\begin{align}
		C_{02g}=0.
		\label{Eq.16}
	\end{align}
    Based on Eq. (\ref{Eq.16}), the optimal conditions for universal photon blockade of right cavity can be derived as follows
    \begin{align}
		&\Delta_c =0 \text{,} \nonumber \\
        &\varphi =2k\pi \text{,}\text{ }\text{ }\text{ }\text{ }\text{ }
        \text{ }\text{ }\text{ }\text{ }\text{ }k= 0,\pm 1,\pm 2,\dotsc\nonumber \\
        &g=|2r-1|\sqrt{\frac{\gamma\kappa}{8r(2-r)}}.
		\label{Eq.17}
	\end{align}

In deriving the aforementioned optimal conditions, we have fully incorporated both conventional and unconventional mechanisms \cite{30}.

\begin{figure}
\begin{minipage}{0.5\textwidth}

		\centering
		\subfigure{\includegraphics[width=6cm]{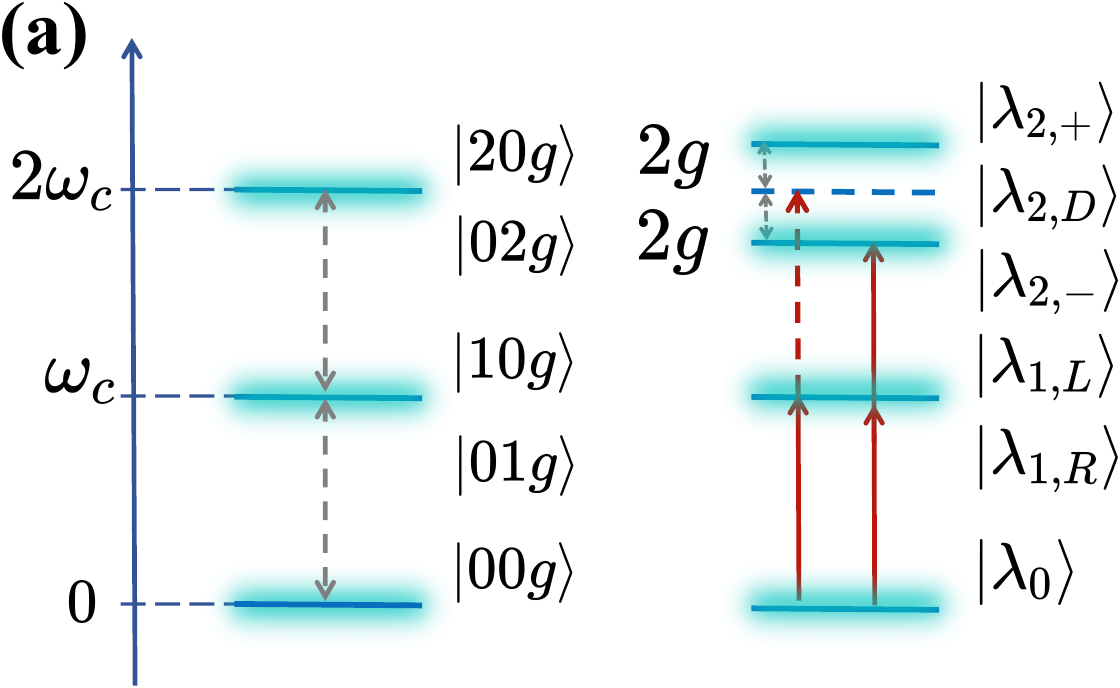}}
        \hspace{0.01cm}
        \subfigure{\includegraphics[width=9cm]{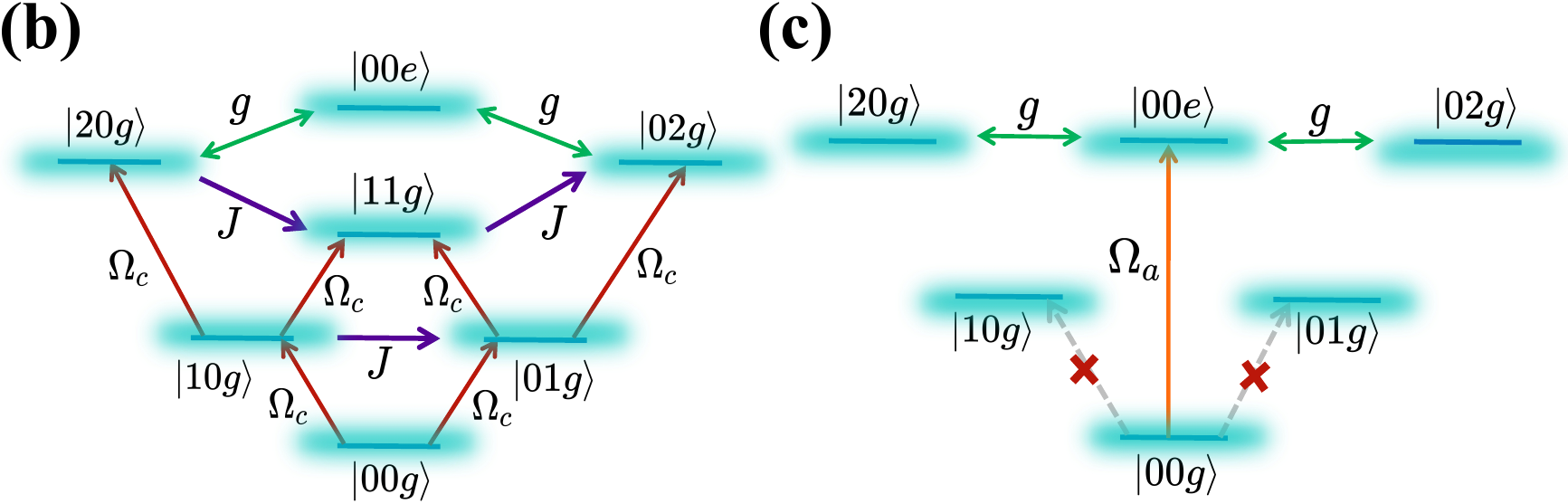}}

		\caption{(a) Energy spectrum of the two-photon two-mode JC model. (b) Transition pathways of the cavity drive in the quantum interference model. In the panels, $J=-i\kappa|r|e^{i\varphi}$. (c) Transition pathways of the atom drive in the quantum interference model. }
		\label{fig2}
        \end{minipage}
	\end{figure}

The emergence of universal photon blockade is mechanistically driven by a dual-origin: the single-photon resonance effect and the pathways of destructive quantum interference. To elucidate the single-photon resonance mechanism, we analyze the eigenstates and the corresponding energy spectrum of the two-photon two-mode JC model, as depicted in Fig. \ref{fig2}(a). The ground state of the system can be denoted as $|\lambda_{0}\rangle=|00g\rangle$ with the eigenvalue $E_0 = 0$. The first excited states of the left and right cavity can be expressed as $|\lambda_{1,L}\rangle=|10g\rangle$ and $|\lambda_{1,R}\rangle=|01g\rangle$ with the same eigenvalue $E_{1}= \omega_c$. In the two-excitation subspace, the state $\vert{}11g\rangle$ is strictly decoupled from the nonlinear evolution \cite{61}, because the intra-cavity two-photon interactions ($c_L^2$ and $c_R^2$) strictly prohibit any transition between $\vert{}11g\rangle$ and the atomic excited state $\vert{}00e\rangle$. Consequently, this state does not participate in Rabi
splitting or coherent interference, and can be safely treated
as a spectator state in the system dynamics. \textcolor{black}{ By solving the energy eigenvalue equation, the eigenvalues of the system can be expressed as $E_2 = 2\omega_c,2\omega_c\pm 2g$ in the two-excitation subspace and the corresponding eigenstate can be expressed as}
\begin{subequations}
\begin{align}
		&|\lambda_{2,D}\rangle = \frac{1}{\sqrt{2}}(|20g\rangle-|02g\rangle), \label{18,a}\\
        &|\lambda_{2,+}\rangle = \frac{1}{\sqrt{2}}|00e\rangle+\frac{1}{2}(|20g\rangle+|02g\rangle), \label{18,b}\\
        &|\lambda_{2,-}\rangle = -\frac{1}{\sqrt{2}}|00e\rangle+\frac{1}{2}(|20g\rangle+|02g\rangle),
		\label{18,c}
	\end{align}
\end{subequations}
where $|\lambda_{2,D}\rangle$ is a dark state in the system. Under the resonance condition $\omega_d = \omega_c$, the single-photon state $|\lambda_{1,s}
\rangle$ ($s=L,R$) population is substantially enhanced via resonant absorption. Meanwhile, the  energy-level gap arising from the  coupling strength $g$ effectively suppresses the excitation of two-photon states $|\lambda_{2,\pm}\rangle$, leading to a negligible population.
    
The schematic representation of destructive quantum interference is depicted in Fig. \ref{fig2}(b). There are multiple different pathways for the system to reach the two-photon state in the right cavity $|02g\rangle$, which mainly include (i) $|00g\rangle \xrightarrow{\Omega_c} |01g\rangle \xrightarrow{\Omega_c} |02g\rangle$, (ii) $|00g\rangle \xrightarrow{\Omega_c} |10g\rangle \xrightarrow{J} |01g\rangle \xrightarrow{\Omega_c}|02g\rangle $, (iii) $|00g\rangle \xrightarrow{\Omega_c} |10g\rangle \xrightarrow{\Omega_c} |11g\rangle \xrightarrow{J}|02g\rangle $, and (iv) $|00g\rangle \xrightarrow{\Omega_c} |10g\rangle \xrightarrow{\Omega_c} |20g\rangle \xrightarrow{g}|00e\rangle \xrightarrow{g}|02g\rangle $. The mutual interference among these excitation pathways suppresses the population of the state $|02g\rangle$. 
 \begin{figure}
		\begin{minipage}{0.5\textwidth}
			\centering
		\subfigure{\includegraphics[width=4.3cm]{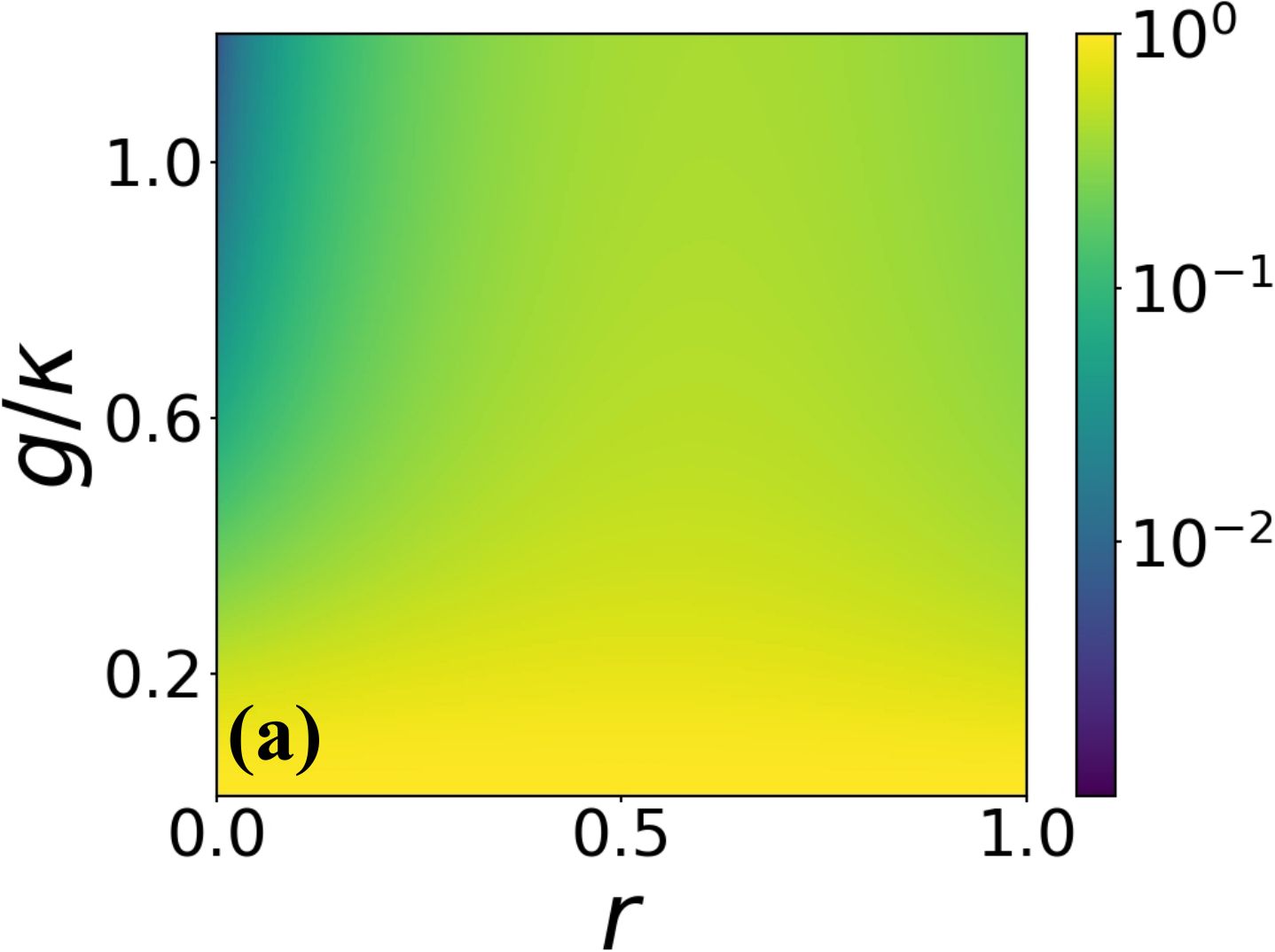}}
		\hspace{0.01cm}
		\subfigure{\includegraphics[width=4.3cm]{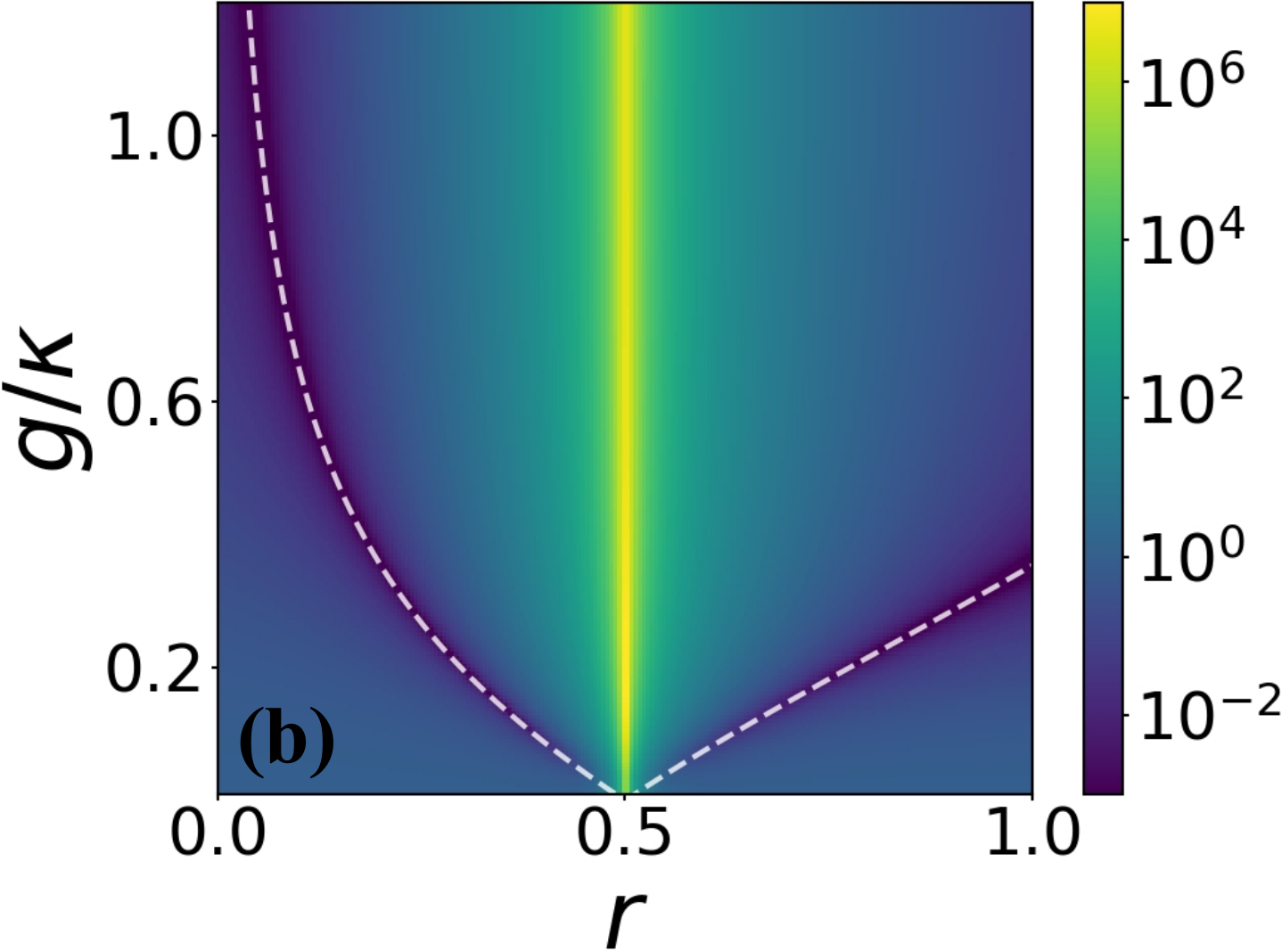}}

		\end{minipage}\hfill
		\caption {Density plots of the steady-state equal-time second-order correlation functions versus the nonlinear atom-cavity coupling strength $g/\kappa$ and the mirror reflectivity $|r|$ under cavity driving. (a) The correlation function in the left cavity, $g_L^{(2)}(0)$. (b) The correlation function in the right cavity, $g_R^{(2)}(0)$. The system parameters are chosen as $\gamma/\kappa=1$, $\Omega_c/\kappa=0.01$, $\Delta_a = 0$, and $\varphi=0$ and the white dashed line in the figure corresponds to the relation: $g/\kappa=|2r-1|\sqrt{{1}/{8r(2-r)}}$.}
		\label{fig3}
	\end{figure}
\begin{figure*}
		\begin{minipage}{1.0\textwidth}
			\centering
		\subfigure{\includegraphics[width=5.3cm]{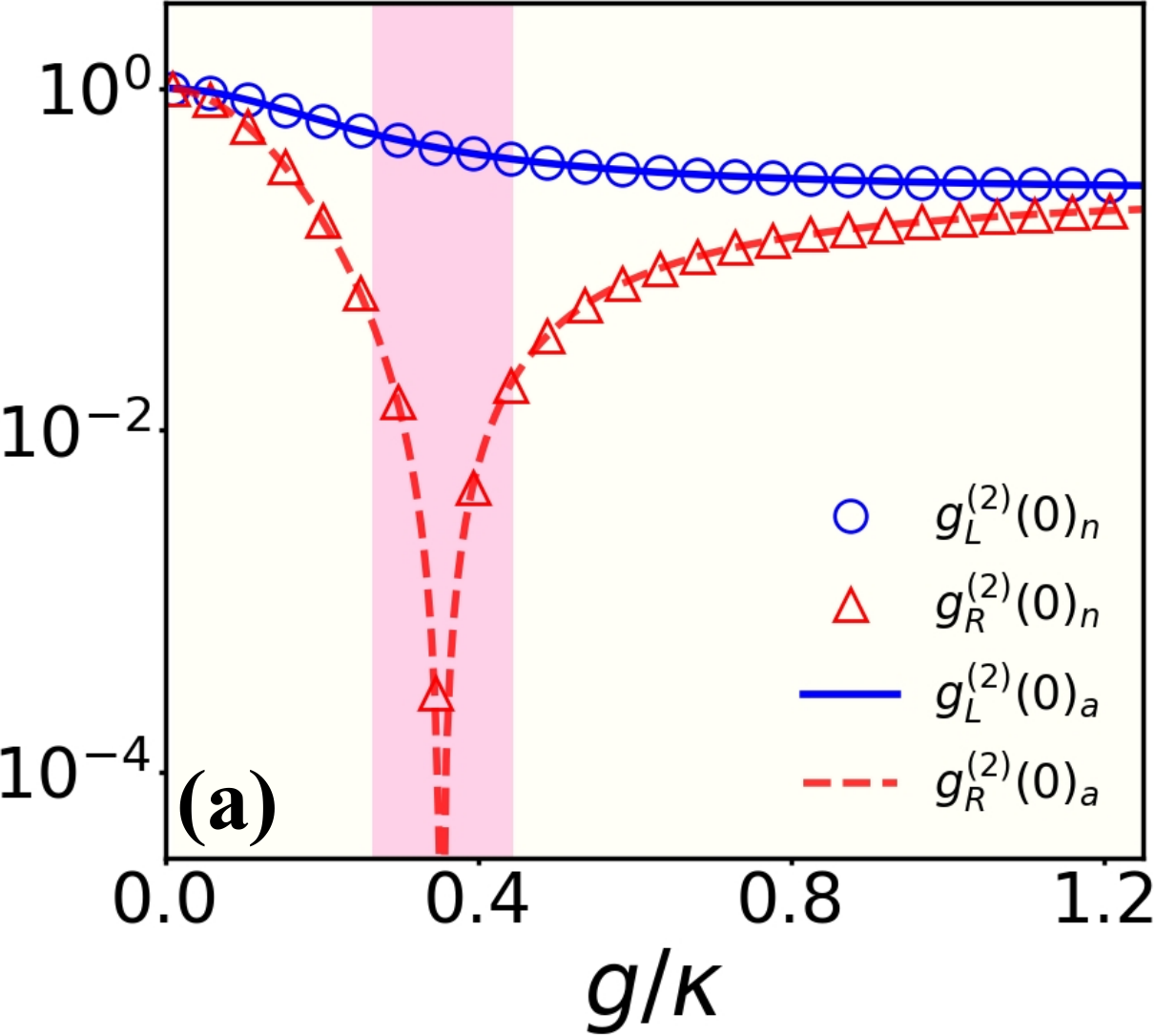}}
		\hspace{0.01cm}
		\subfigure{\includegraphics[width=5.3cm]{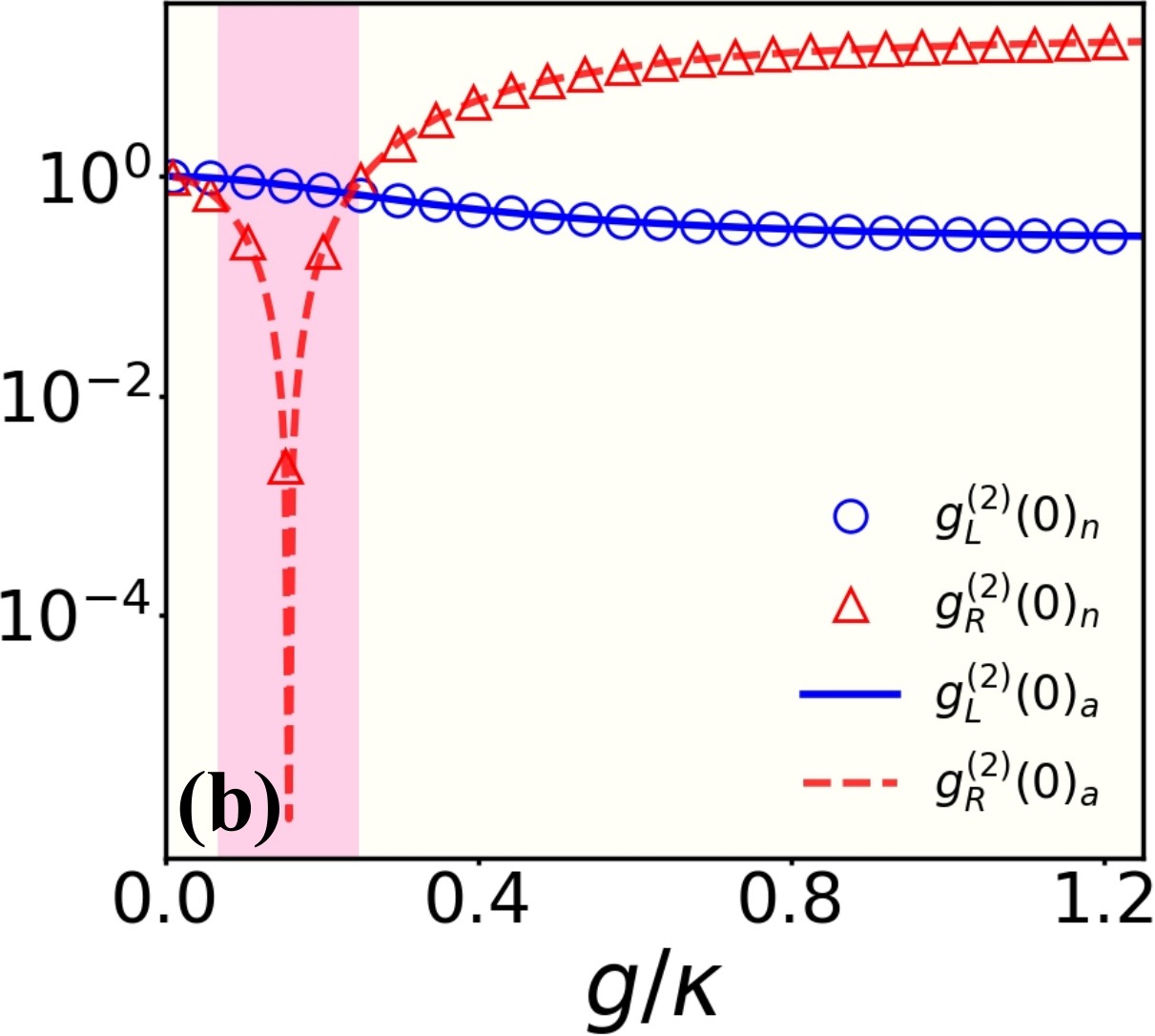}}
        \hspace{0.01cm}
        \subfigure{\includegraphics[width=5.3cm]{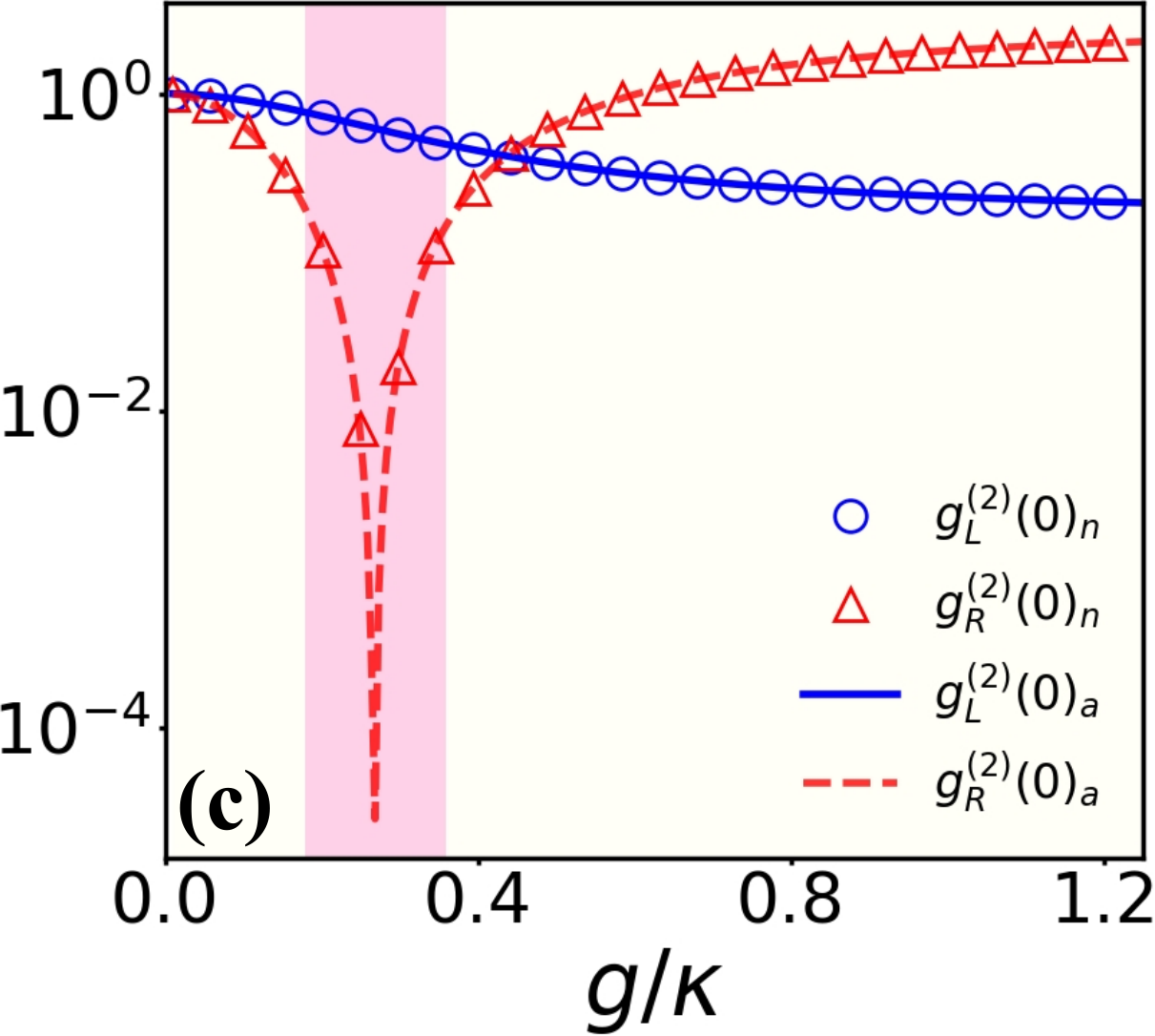}}
		\\
		\subfigure{\includegraphics[width=5.3cm]{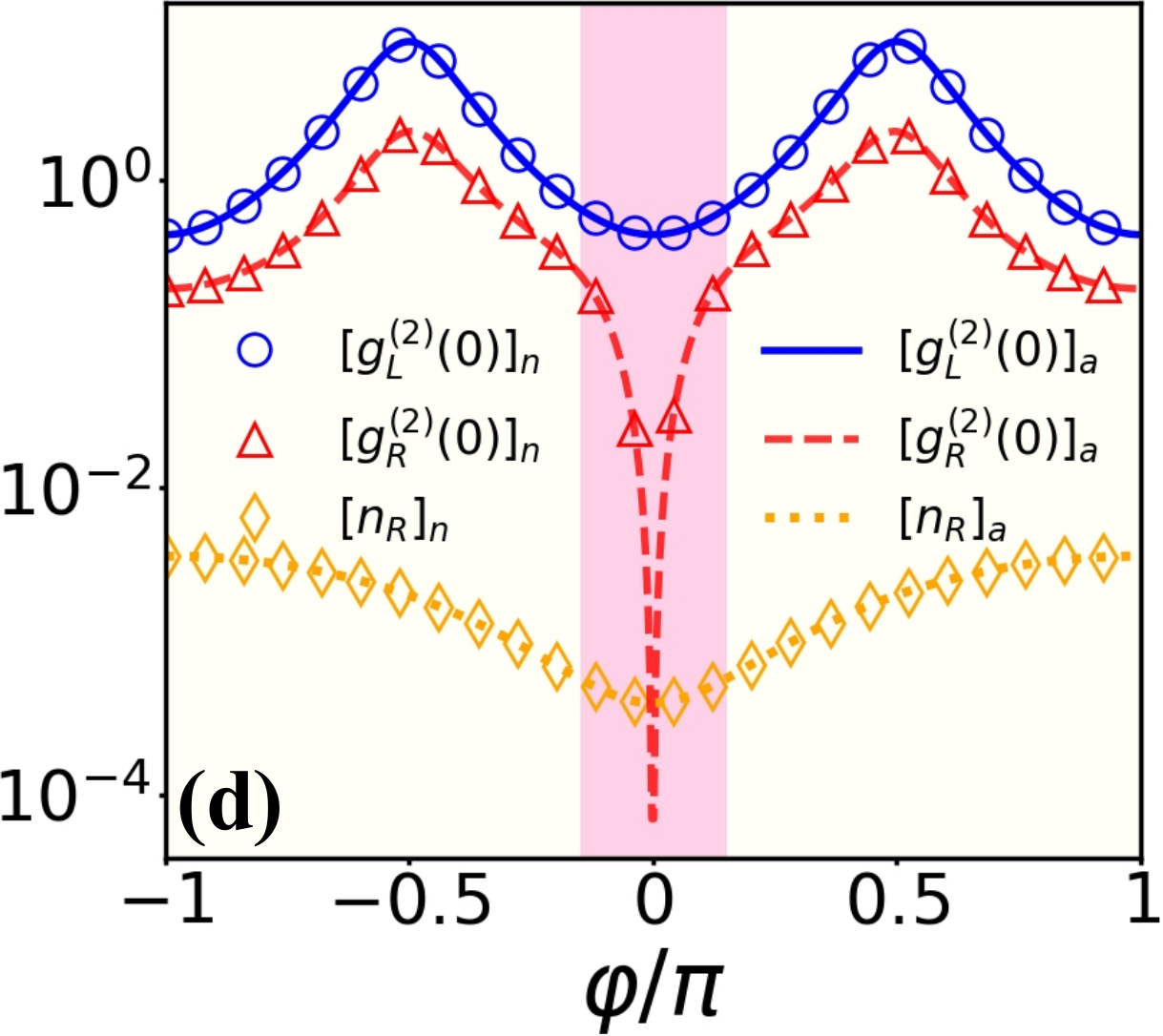}}
		\hspace{0.01cm}
		\subfigure{\includegraphics[width=5.3cm]{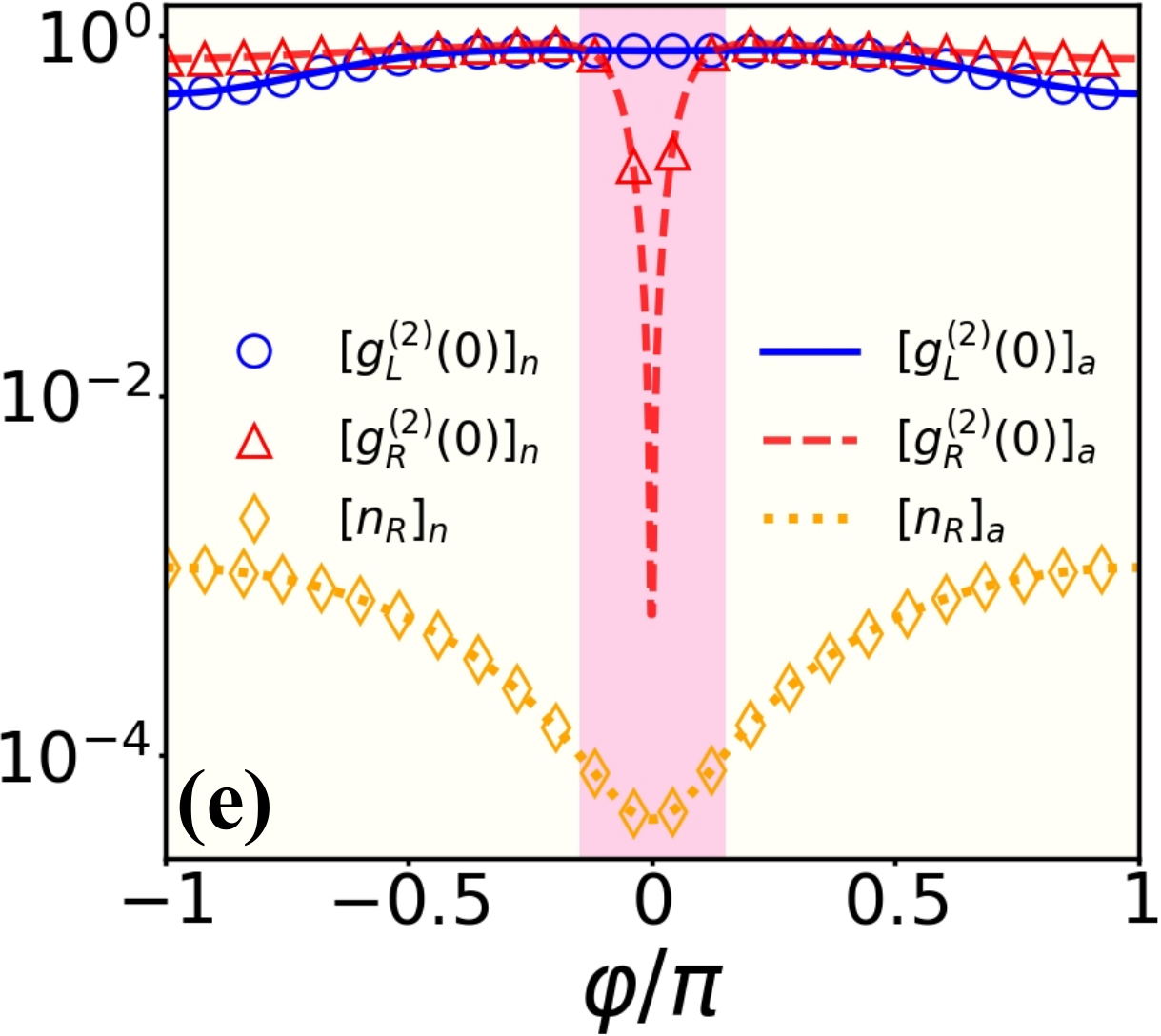}}
        \hspace{0.01cm}
        \subfigure{\includegraphics[width=5.3cm]{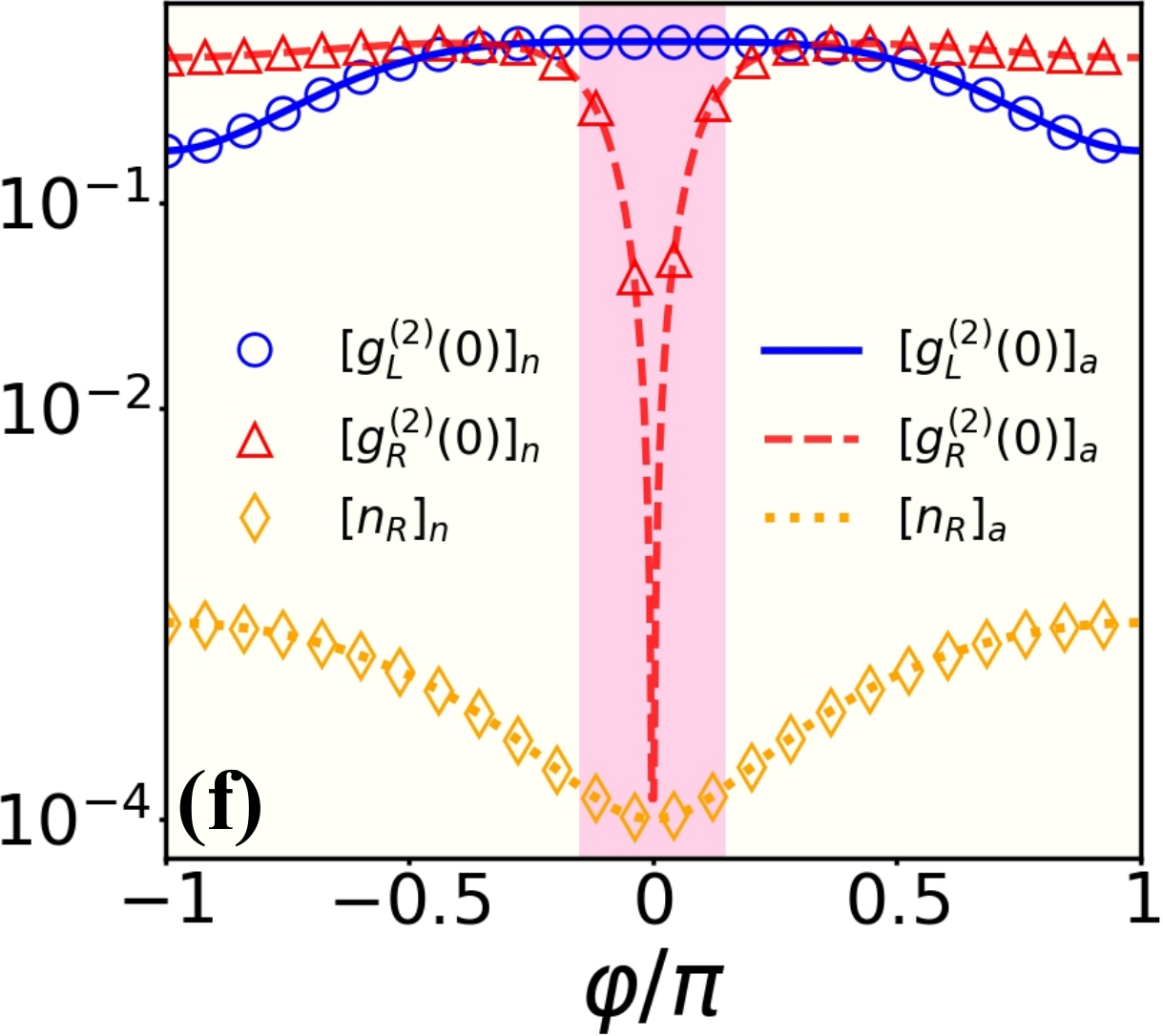}}
        \\
        \subfigure{\includegraphics[width=5.3cm]{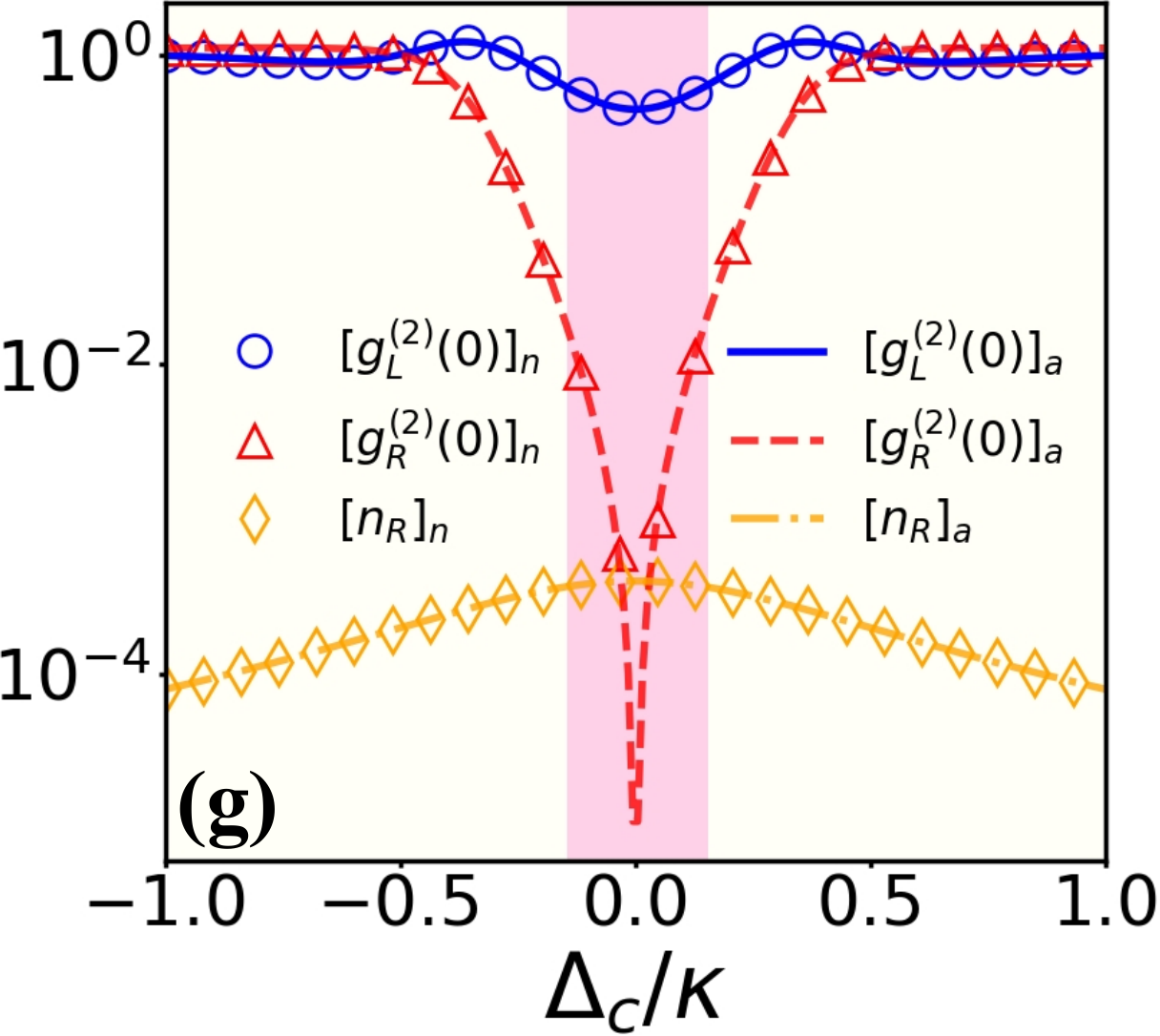}}
		\hspace{0.01cm}
		\subfigure{\includegraphics[width=5.3cm]{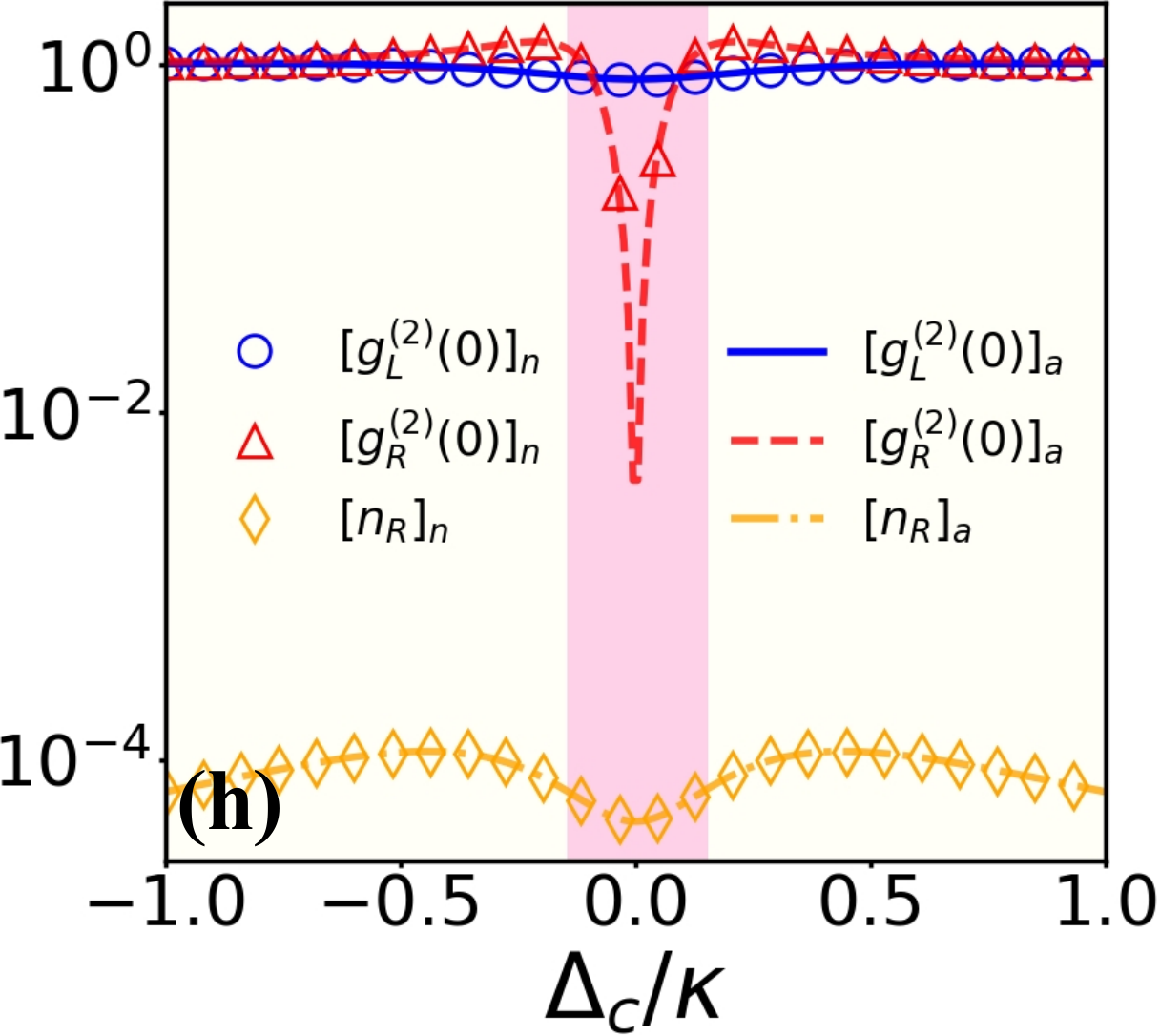}}
        \hspace{0.01cm}
        \subfigure{\includegraphics[width=5.3cm]{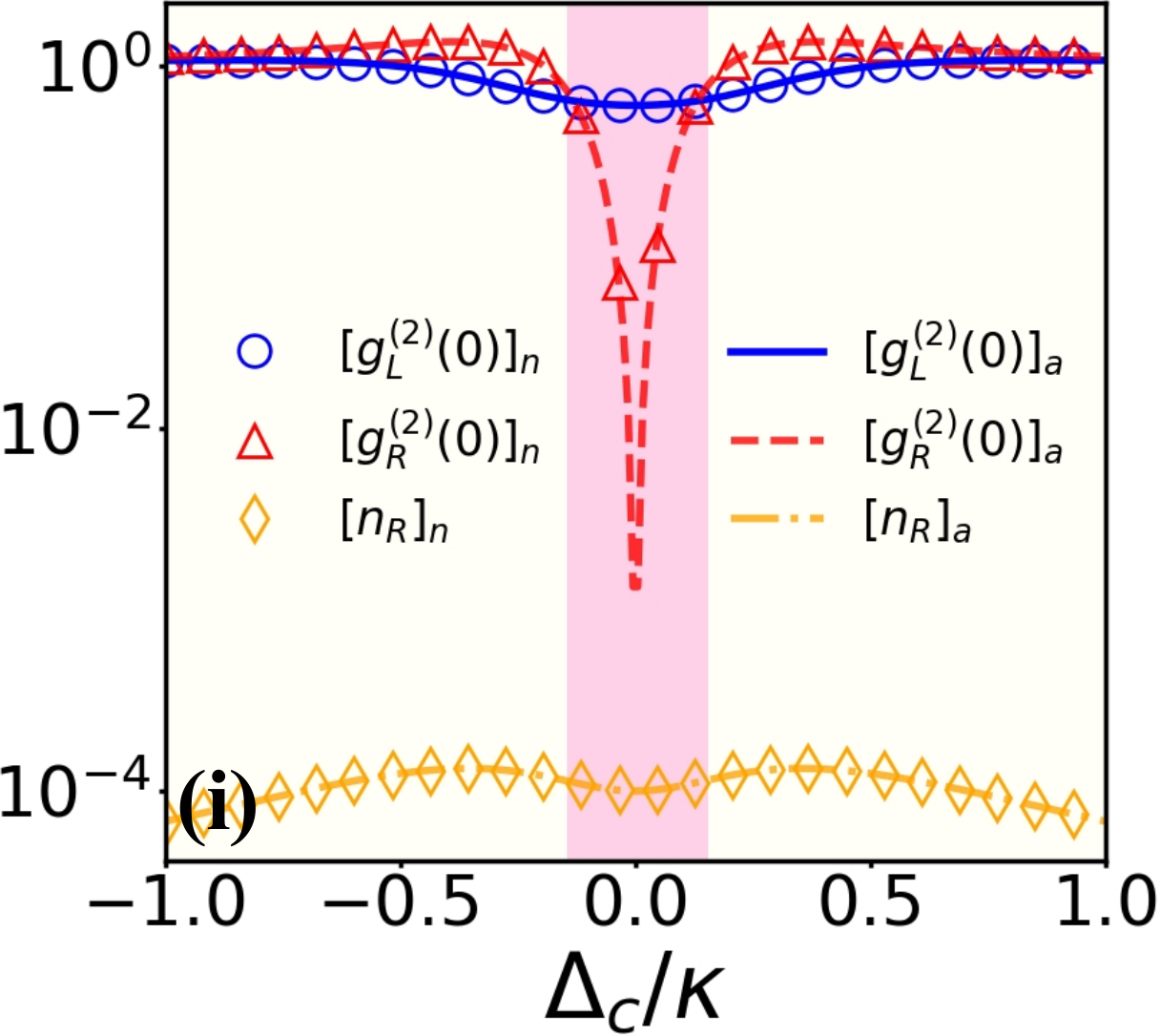}}
		\end{minipage}\hfill
		\caption{ \textcolor{black}{(a)-(c) $g^{(2)}_{s}(0)$ as a function of $g$ in the resonant case $\Delta_c=0$ for $\Omega_c/\kappa=0.01$, $\gamma/\kappa=1$ and (a) $|r|=1$,(b) $|r|=0.33$ and (c) $|r|=0.25$ }. (d)-(f) $g^{(2)}_{s}(0)$ and $n_R$ as a function of $\varphi$ in the resonant case $\Delta_c=0$ for $\Omega_c/\kappa=0.01$, $\gamma/\kappa=1$ and (d) $|r|=1,g/\kappa=0.354$,(e) $|r|=0.33,g/\kappa=0.158$ and (f) $|r|=0.25,g/\kappa = 0.267$. (g)-(i) $g^{(2)}_{s}(0)$ and $n_R$ as a function of $\Delta_c$  for $\Omega_c/\kappa=0.01$, $\gamma/\kappa=1$, $\varphi=0$ and (g) $|r|=1,g/\kappa=0.354$, (h) $|r|=0.33,g/\kappa=0.158$ and (i) $|r|=0.25,g/\kappa = 0.267$. In the panels, the subscripts "$n$" and "$a$" denote numerical and analytical solutions, respectively.}
		\label{fig4}
	\end{figure*}
   
Regarding the left cavity, the presence of different transition pathways to $|20g\rangle$ can also be observed, which can be expressed as (i) $|00g\rangle \xrightarrow{\Omega_c} |10g\rangle \xrightarrow{\Omega_c} |20g\rangle$, (ii)  $|00g\rangle \xrightarrow{\Omega_c} |01g\rangle \xrightarrow{\Omega_c} |02g\rangle \xrightarrow{g}|00e\rangle \xrightarrow{g}|20g\rangle $, and (iii) $|00g\rangle \xrightarrow{\Omega_c} |10g\rangle \xrightarrow{J} |01g\rangle\xrightarrow{\Omega_c}|02g\rangle\xrightarrow{g}|00e\rangle\xrightarrow{g}|20g\rangle$. Thus, the quantum interference  manifests.

\begin{figure*}
		\begin{minipage}{1\textwidth}
			\centering
		\subfigure{\includegraphics[width=5.4cm]{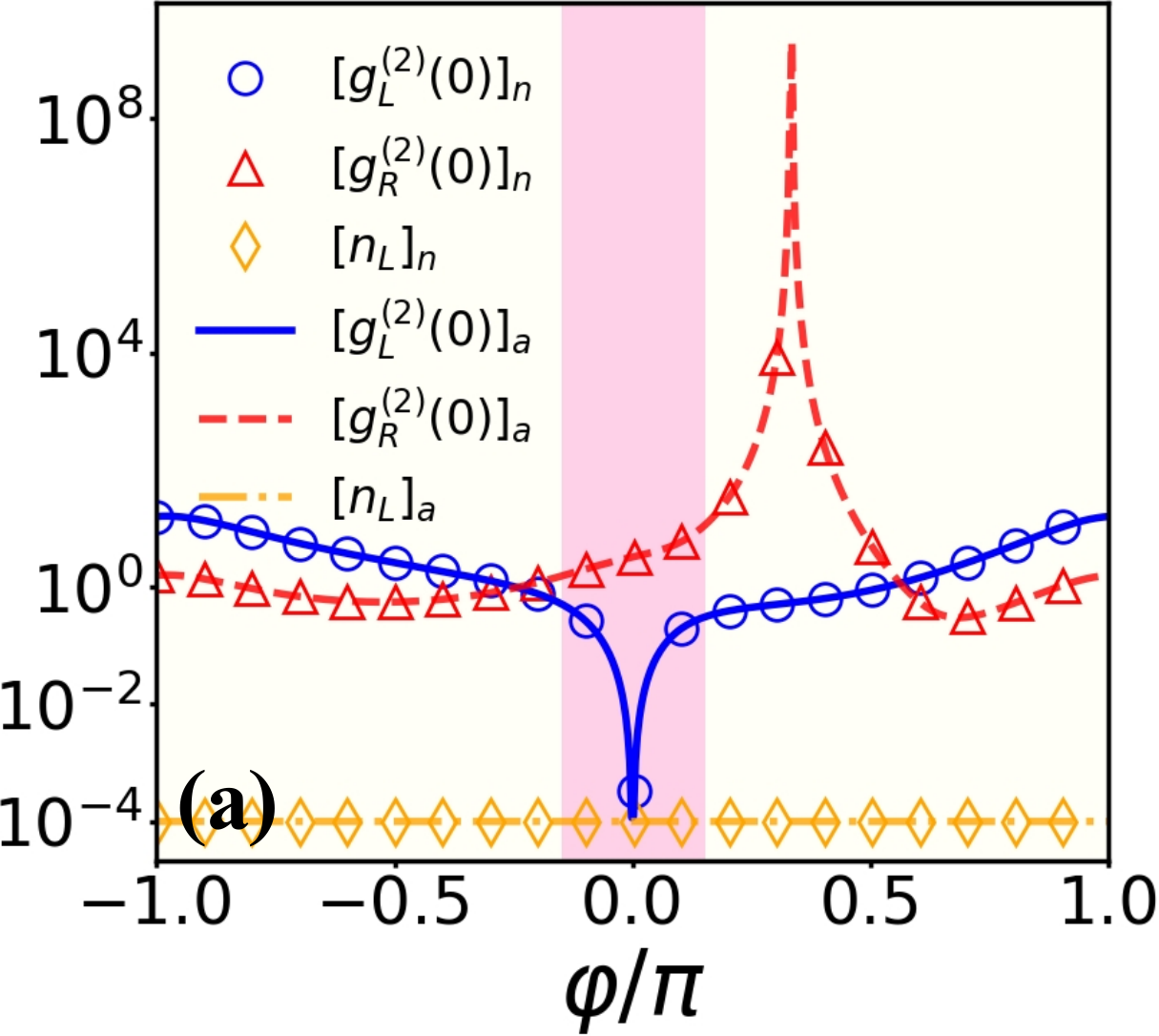}}
        \hspace{0.01cm}
		\subfigure{\includegraphics[width=5.4cm]{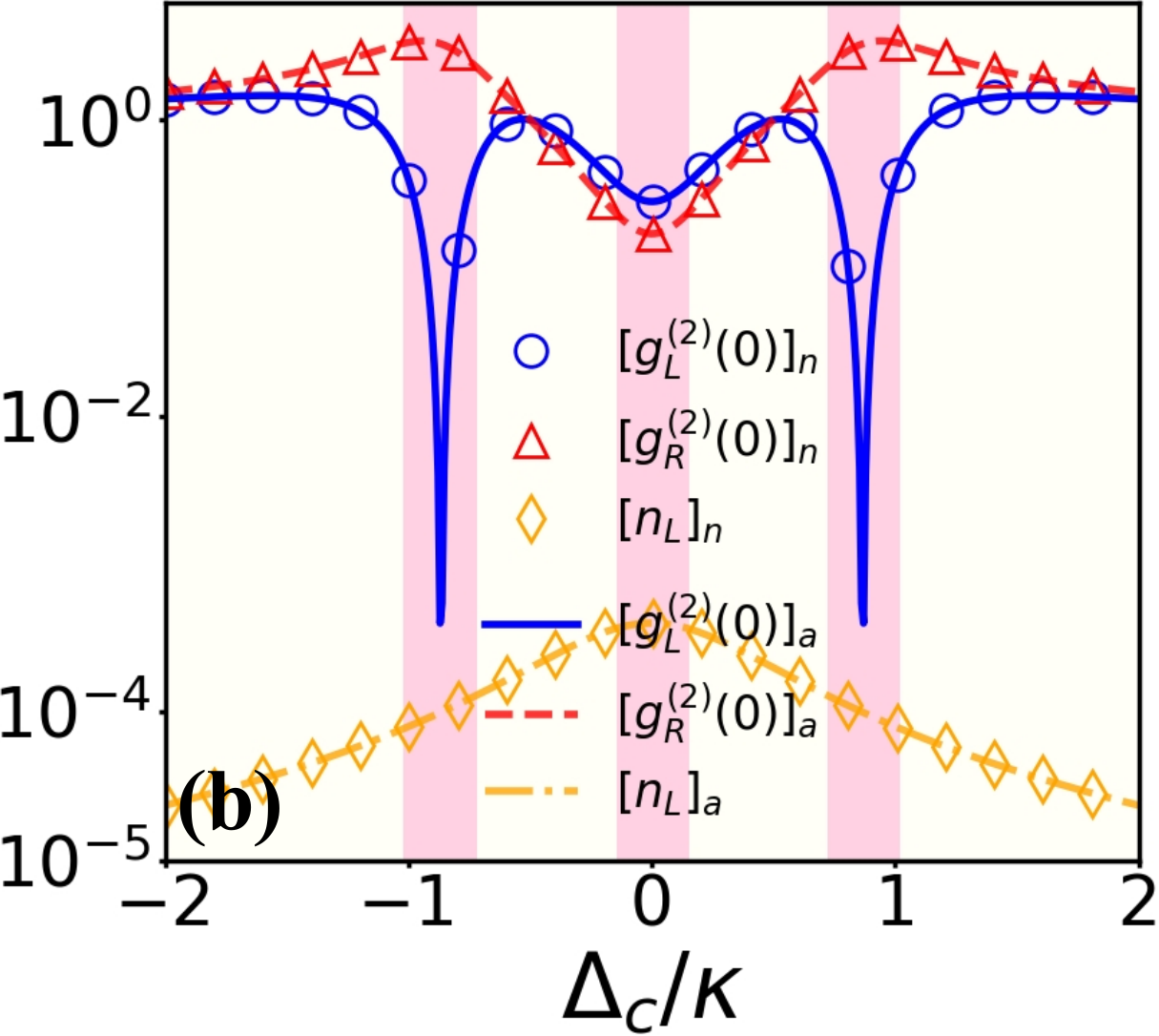}}
        \hspace{0.01cm}
        \subfigure{\includegraphics[width=5.4cm]{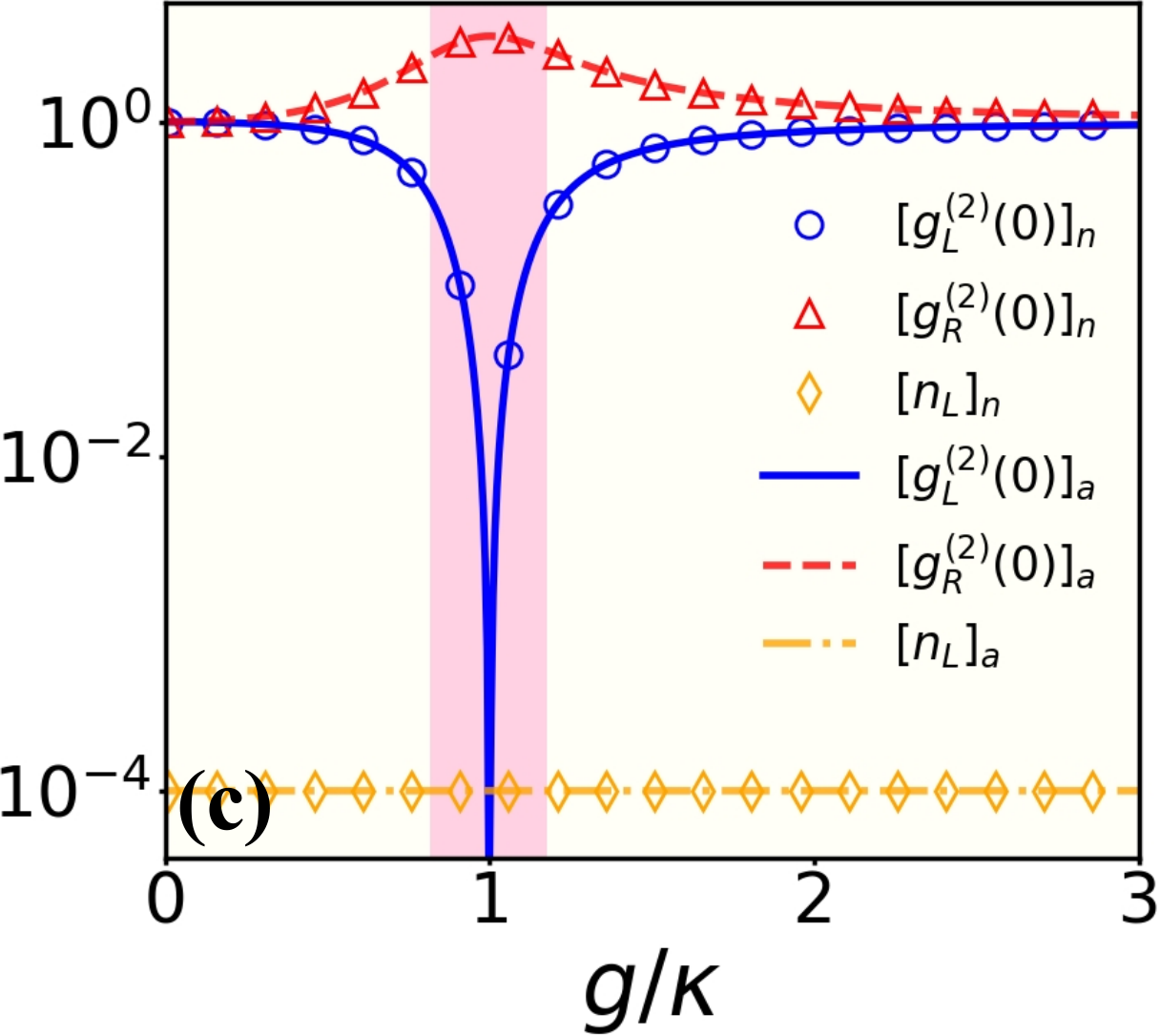}}
		
		\end{minipage}\hfill
		\caption{ \textcolor{black}{(a) Plots of the $g_{s}^{(2)}(0)$($s=L,R$) and $n_L$ as a function of $\varphi$ with system parameters $|r|=1$, $\gamma/\kappa=1$, $\Delta_c/\kappa=0.866$, $g/\kappa=1$. (b) Plots of the $g_{s}^{(2)}(0)$($s=L,R$) and $n_L$ as a function of $\Delta_c$ with system parameters $|r|=1$, $\varphi=0$ , $\gamma/\kappa=1$, $g/\kappa=1$. (c) Plots of the $g_{s}^{(2)}(0)$($s=L,R$) and $n_L$ as a function of $\Delta_c$ with system parameters $|r|=1$, $\varphi=0$, $\gamma/\kappa=1$, $\Delta_c/\kappa=0.866$. In the panels, the subscripts "$n$" and "$a$" denote numerical and analytical solutions, respectively.}}
		\label{fig5}
	\end{figure*}
To validate our analytical results, we numerically solve Eq. (\ref{Eq.4}) by replacing $H_t$ with $H_{\rm sys}$. In Fig. \ref{fig3}, we present $g^{(2)}_s(0)$ ($s=L,R$) as a function of the atom-cavity coupling strength $g$ and $|r|$ with atomic decay rates $\gamma/\kappa=1$. $\Omega_c/\kappa=0.01$ is set for the weak driving condition, and the parameters $\Delta_c=0$ and $\varphi=0$ are set to meet the first two of the optimal conditions. As illustrated in Fig. \ref{fig3}(a), $g_L^{(2)}(0)$ constantly hovers around 1 for the left cavity. This indicates the absence of an obvious PB, aligning perfectly with our analytical finding that the exact condition for complete destructive interference ($C_{20g}=0$) cannot be satisfied. However, the photon statistical properties of the right cavity in Fig. \ref{fig3}(b) are profoundly different. As dictated by the optimal condition in Eq. (\ref{Eq.17}), the correlation function $g_R^{(2)}(0)$ drops abruptly toward zero along the white dashed line. This conclusively demonstrates that by precisely matching the parameters $g/\kappa$ and $|r|$, the interaction between different excitation pathways effectively suppresses two-photon excitation. Consequently, a strong PB can be achieved regardless of whether $g < \kappa$, $g \sim \kappa$, or $g > \kappa$. Furthermore, it is noteworthy in Fig. \ref{fig3}(b) that the value of $g_R^{(2)}(0)$ surges around $|r|= 0.5$, exhibiting a prominent photon bunching effect. Previously, we only focused on the excitation pathways leading to the state $\vert{}02g\rangle$, and now we analyze the transitions to the single-photon state $\vert{}01g\rangle$. As can be seen from Fig. \ref{fig2}(b), there exist two distinct excitation pathways to the state $\vert{}01g\rangle$, which can be represented as (i) $\vert{}00g\rangle  \xrightarrow{\Omega_c} \vert{}01g\rangle$ and (ii) $\vert{}00g\rangle  \xrightarrow{\Omega_c} \vert{}10g\rangle  \xrightarrow{J} \vert{}01g\rangle$. The interaction between these two pathways leads to destructive quantum interference, which suppresses the population of the state $\vert{}01g\rangle$. The condition for perfect destructive interference can be determined by $C_{01g} = 0$. Under the resonance condition, setting $\varphi = 0$ yields $|r| = 0.5$. Therefore, it becomes clear that at $|r| = 0.5$, the system completely suppresses the excitation of the state $\vert{}01g\rangle$, but does not perfectly suppress the excitation of the state $\vert{}02g\rangle$. According to  $g_R^{(2)}(0)\approx2|C_{02g}|^{2}/|C_{01g}|^{4}$, as $C_{01g} \to 0$, $g_R^{(2)}(0)$ approaches infinity.

We now investigate the behavior of $g_{ s}^{(2)}(0)$ ($s=L,R$) within the coupling $g/\kappa$ and the phase $\varphi$. Figs. \ref{fig4}(a)–(c) plot the $g_{ s}^{(2)}(0)$ ($s=L,R$) for identical driving strength $\Omega_c/\kappa = 0.01$ and various mirror reflectivities $|r|=1,0.33$ and $0.25$. It is observed that the photon statistics of the left cavity exhibit remarkable robustness to both the coupling strength $g$ and reflectivity $|r|$ at resonance; the corresponding correlation function stays close to 1, characteristic of a Poissonian distribution. Conversely, the right cavity displays a correlation function that drops toward zero  within regime $g/\kappa <1$, which serves as a signature of PB.

Figs. \ref{fig4}(d)–(f) illustrate the dependence of the correlation function $g^{(2)}(0)$ and mean photon number of the right cavity $n_R$ on the phase $\varphi$. The parameters are chosen as $|r| = \{1, 0.33, 0.25\}$, with the corresponding coupling strengths $g/\kappa = \{0.354, 0.158, 0.267\}$ computed via Eq. (\ref{Eq.17}), $\gamma/\kappa=1$ and $\Omega_c/\kappa=0.01$. We observe that $g_{ L}^{(2)}(0)>1$ and $g_{ R}^{(2)}(0)>1$ at $|r|=1,\varphi=\pm  \pi/2$, suggesting the occurrence of photon bunching effect both in the left and right cavity. When the phase $\varphi$ approaches zero,  $g_{ R}^{(2)}(0)$  sharply  drops to zero, demonstrating the occurrence of PB.

Furthermore, we investigate the $g^{(2)}(0)$ and $n_R$ as functions of the detuning $\Delta_c$, with the remaining parameters satisfying the optimal conditions. As depicted in Figs. \ref{fig4}(g)-(i), $g_{R}^{(2)}(0)$ is nearly constant but falls precipitously to 0 at $\Delta_c=0$, Simultaneously, $n_R$ reaches its extremum. We observe in Fig. \ref{fig4}(g) that $n_R$ peaks at $\Delta_c = 0$, corresponding to the minimum of $g_{R}^{(2)}(0)$, indicating that CPB occurs, whereas UPB occurs under other parameter conditions.

   \textcolor{black}{\subsection{ Off-resonance case }}
We now examine the photon statistics of the two-cavity system in the off-resonance regime $\Delta_{a}=2\Delta_{c}\ne 0 $. We first consider the left cavity and set $|r|=1$. Similarly, the optimal conditions for  photon blockade are given by 
\begin{align}
		C_{20g}=0.
		\label{Eq.19}
	\end{align}
Consequently, the optimal conditions  are derived as
\begin{align}
		 &\varphi =2k\pi \text{,}\text{ }\text{ }\text{ }\text{ }\text{ }\text{ }\text{ }\text{ }\text{ }\text{ }\text{ }\text{ }\text{ }\text{ }
        k= 0,\pm 1,\pm 2,\dotsc\nonumber \\
      &\Delta_c=\pm\frac{\sqrt{5\gamma\kappa+6\kappa^2+\kappa\varrho}}{4\sqrt{2}}, \nonumber \\
      &g=\frac{\sqrt{5\gamma^2+12\gamma\kappa+20\kappa^2+(\gamma+6\kappa)\varrho}}{8\sqrt{2}},
		\label{Eq.20}
	\end{align}
where $\varrho= \sqrt{25\gamma^2+76\gamma\kappa+68\kappa^2}$. Additionally, the optimal conditions for  photon blockade in the right cavity are not obtainable in analytical form. We also employ numerical simulations to analyze this result.

Figs. \ref{fig5}(a)-(c) illustrates the variations of correlation function
 $g_{s}^{(2)}(0)$ ($s=L,R$) and mean photon number in the left cavity $n_L$  with respect to $\varphi$, $\Delta_c$ and $g$, respectively. In Fig. \ref{fig5}(a), $\Delta_c/\kappa=0.866$, $g/\kappa=1$, and $\gamma/\kappa=1$ are set to  satisfy the latter two conditions of Eq. (\ref{Eq.20}). Within  $\varphi\in[-\pi,\pi] $, the value of the $g_{L}^{(2)}(0)$ drops sharply and approaches zero at $\varphi = 0$, and $n_L$ remains smooth, signifying the occurrence of a significant PB in the left cavity. Notably, the value of $g_R^{(2)}(0)$ spikes to infinity around $\varphi = \pi/3$, signifying the occurrence of a strong photon bunching effect. We now analyze the physics behind this phenomenon. Similarly, for the right cavity, we need to analyze not only the transition pathways to the two-photon state $\vert{}02g\rangle$, but also the excitation of the single-photon state $\vert{}01g\rangle$. By setting $C_{01g} = 0$, one can obtain the equation $(\Delta_c-\kappa|r|sin\varphi)+i(-\kappa/2+\kappa|r|cos\varphi)=0$. Solving this equation yields $\varphi = \pi/3$. Consequently, at this point, the single-photon excitation is completely suppressed while the two-photon excitation remains, which leads to the photon bunching effect and causes $g_R^{(2)}(0)$ to approach infinity. In Fig. \ref{fig5}(b), we set $\varphi=0$ , $g/\kappa=1$, and $\gamma/\kappa=1$. Within the range of $\Delta_c/\kappa\in[-2,2]$, it is found that $g_L^{(2)}(0)\longrightarrow 0$  only when the detuning $\Delta_c/\kappa = \pm0.866$, but $n_{L}$ fails to reach an extreme value, which demonstrates that UPB occurs. However, we find that $n_L$ exhibits a maximum at $\Delta_c = 0$, where the corresponding $g_{L}^{(2)}(0)$ is less than 1, indicating that CPB occurs at this point. Conversely, $g_R^{(2)}(0)$ fluctuates around unity throughout the $\Delta_c$ range, suggesting that the photons in the right cavity exhibit sub-Poissonian statistics or even the bunching effect. In Fig. \ref{fig5}(c),  $\gamma/\kappa=1$, $\varphi=0$, and $\Delta_c/\kappa=0.866$ are satisfied, with $g/\kappa$ treated as a variable ranging from 0 to 2. We clearly observe $g_{L}^{(2)}(0)$ plummets to its minimum value when  $g/\kappa$ reaches 1 and meanwhile $g_{R}^{(2)}(0)>1 $, resulting in a simultaneous PB effect in the left cavity and a photon bunching effect in the right cavity.
\begin{figure}[htbp] 
  \centering
    \includegraphics[width=0.48\linewidth]{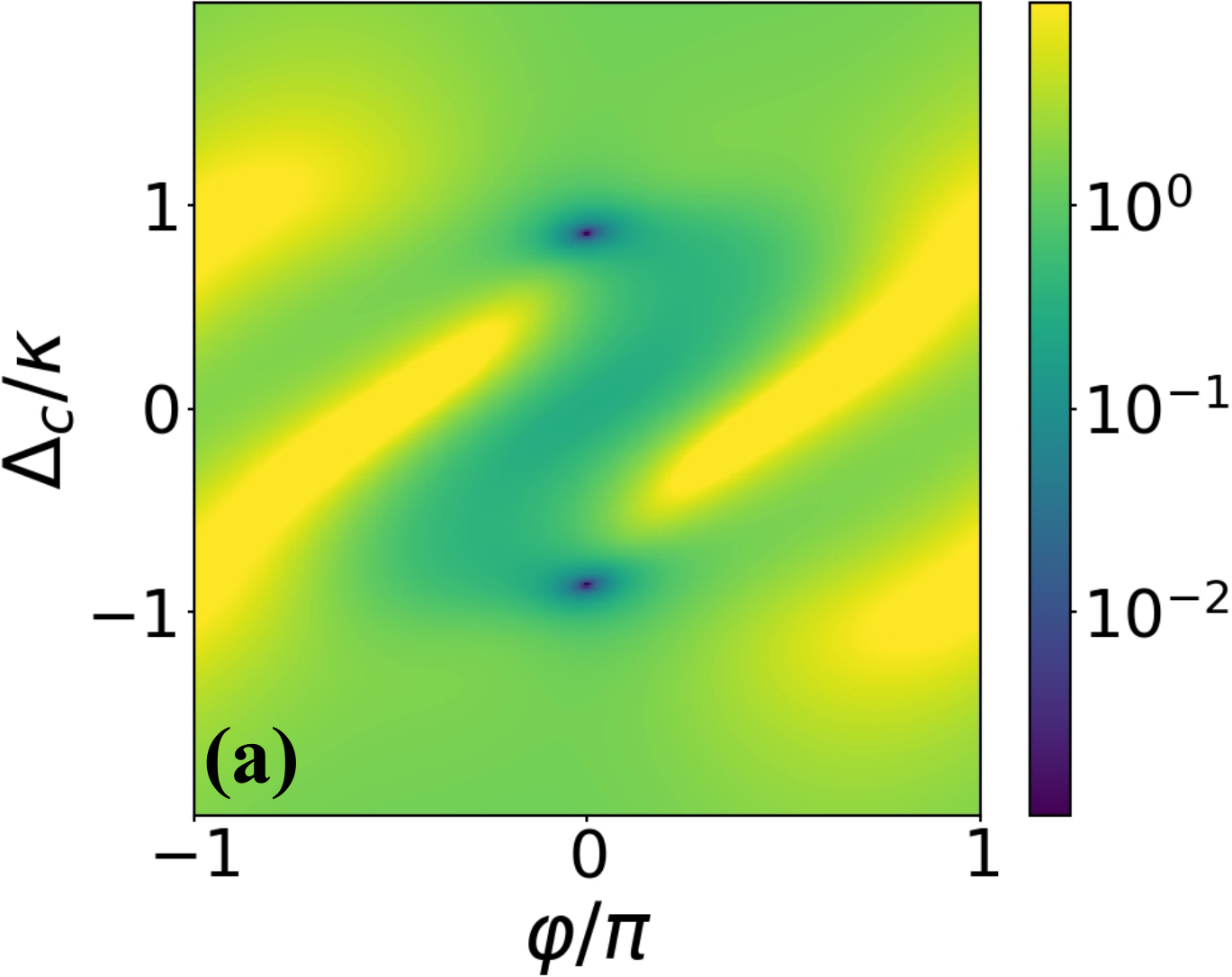}
    \hfill
    \includegraphics[width=0.48\linewidth]{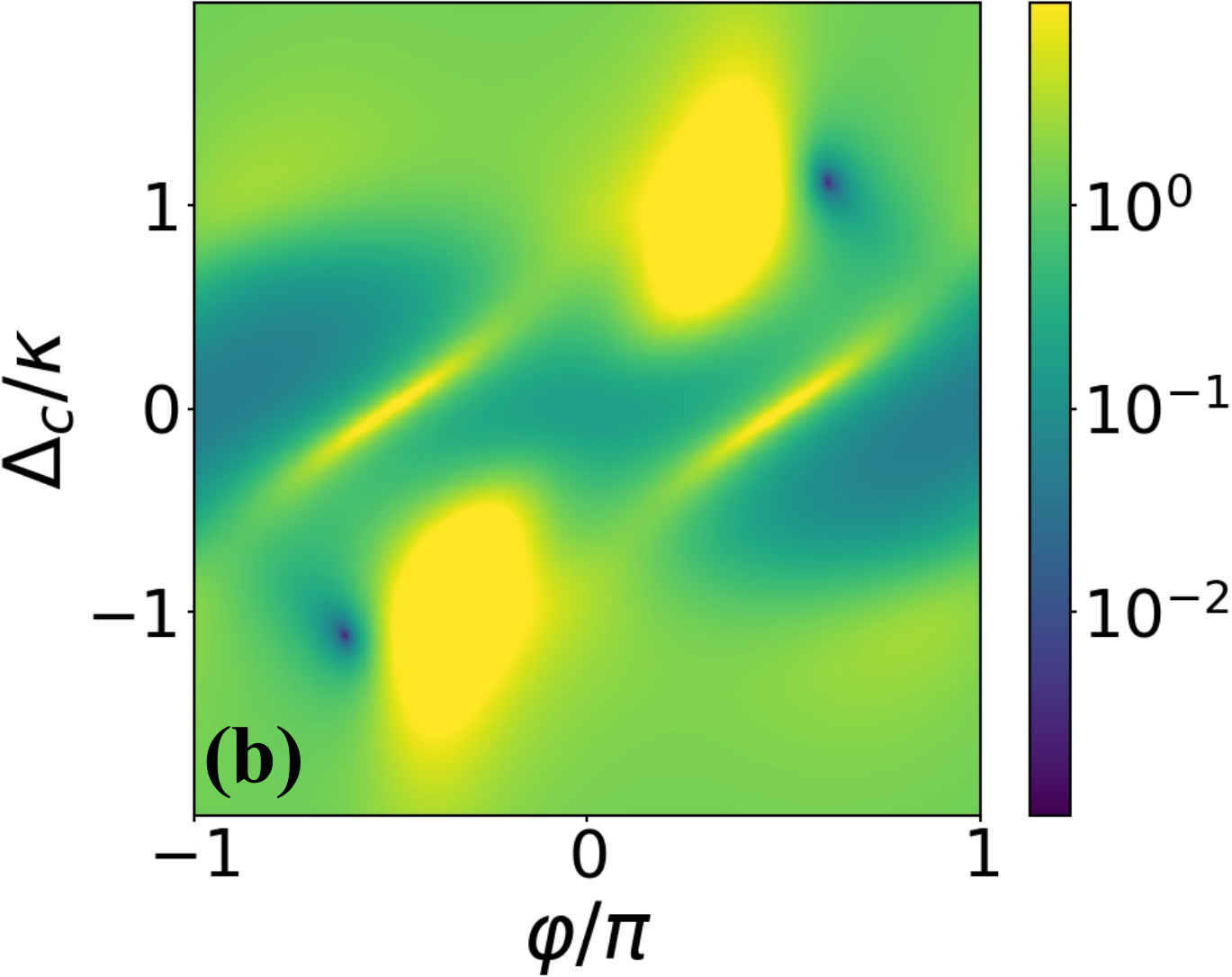}
    
    \vspace{10pt} 
    
    \includegraphics[width=0.48\linewidth]{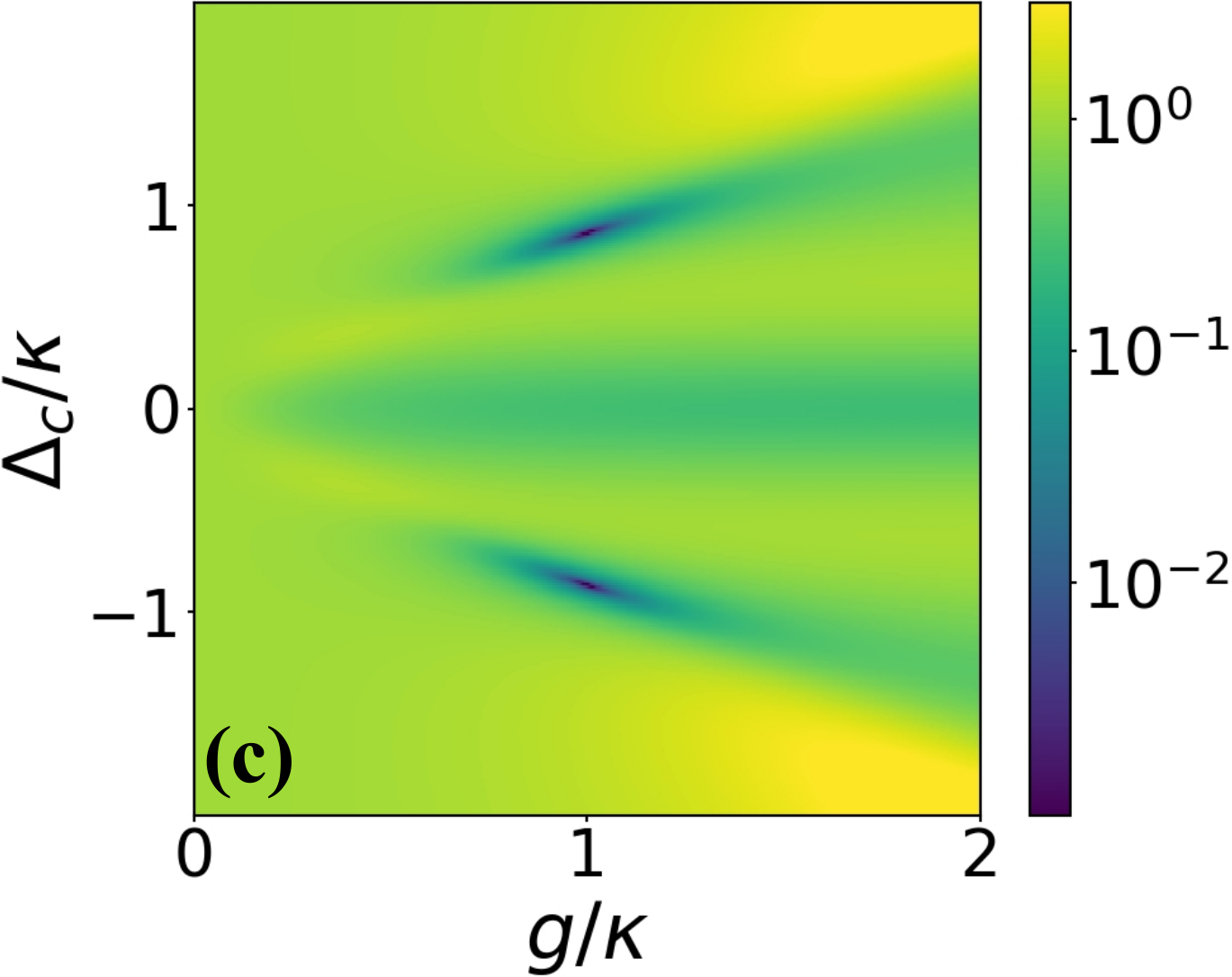}
    \hfill
    \includegraphics[width=0.48\linewidth]{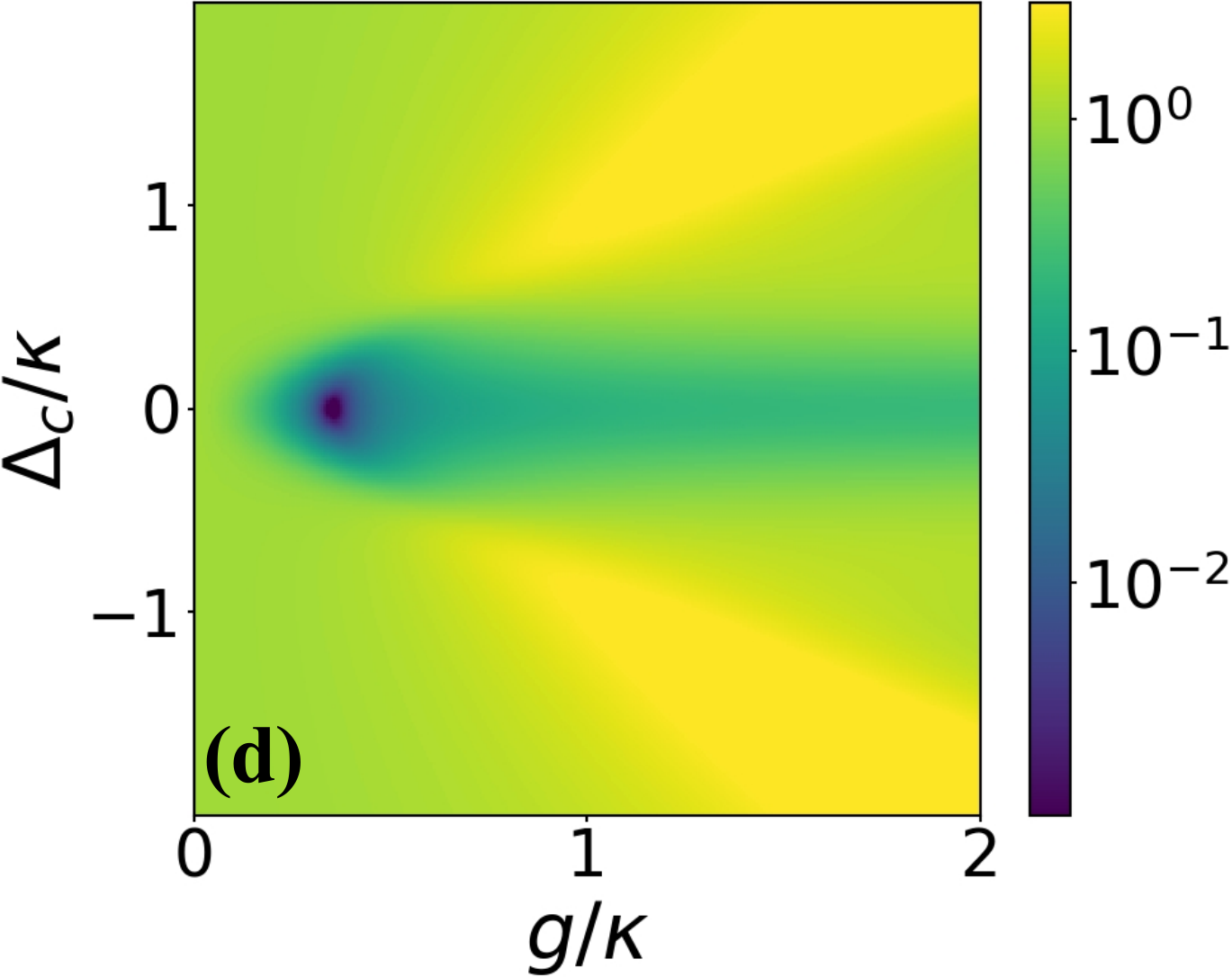}
		\caption{\textcolor{black}{(a) The correlation function $g^{(2)}_L(0)$  as a function of the  $\varphi$ and $\Delta_c/\kappa$ with $\Omega_c/\kappa=0.01$,$\gamma/\kappa = 1$, and $g/\kappa=1$. (b) The correlation function $g^{(2)}_R(0)$  as a function of the  $\varphi$ and the  $\Delta_c/\kappa$, with the same parameter settings as in (a). (c) The correlation function $g^{(2)}_L(0)$  as a function of the coupling strength $g/\kappa$ and the  $\Delta_c/\kappa$ with $\Omega_c/\kappa=0.01$,$\gamma/\kappa = 1$, and $\varphi=0$. (d) The correlation function $g^{(2)}_R(0)$ as a function of the coupling strength $g/\kappa$ and the  $\Delta_c/\kappa$, the parameter settings are the same as (c).}}
		\label{fig6}
	\end{figure}
To further investigate the effects of the three parameters $\varphi$, $\Delta_c$ and $g$ in Eq. (\ref{Eq.20}) on $g_{ s}^{(2)}(0)$ for $\gamma/\kappa=1$, Figs. \ref{fig6}(a)-(b) show $g_{L}^{(2)}(0)$ and $g_{R}^{(2)}(0)$, respectively, as a function of $\varphi$ and $\Delta_c$ with $\Omega_c/\kappa=0.01$, $g/\kappa=1$. It is evident that $g_{L}^{(2)}(0)$  attains its minimum value if and only if $\varphi = 0$ and $\Delta_c/\kappa = \pm0.866$, whereas $g_R^{(2)}(0)$ attains its minima in different parameter regions. In Figs. \ref{fig6}(c)-(d), we fix $\varphi = 0$ and plot $g_{L}^{(2)}(0)$ and $g_{R}^{(2)}(0)$ as functions of $g$ and $\Delta_c$ for $\Omega_c/\kappa = 0.01$. It is clearly observed that the parameter conditions required to achieve the minimum value of $g_{L}^{(2)}(0)$ are $g/\kappa =1$ and  $\Delta_c/\kappa=\pm 0.866$, and that $g_{R}^{(2)}(0)$ reaches its minimum only when $\Delta_c/\kappa = 0$ and $g/\kappa = 0.354$. The numerical simulations align well with our analytical results, showing that the  photon blockade in the left cavity is contingent upon Eq. (\ref{Eq.20}). Specifically, Fig. \ref{fig6}(d) verifies our result of the previous section that universal photon blockade in the right cavity is restricted to the resonance case.

\vspace{3ex} 
\subsection{Robustness analysis}

\begin{figure}[htbp]
    \centering
    \includegraphics[width=0.5\linewidth]{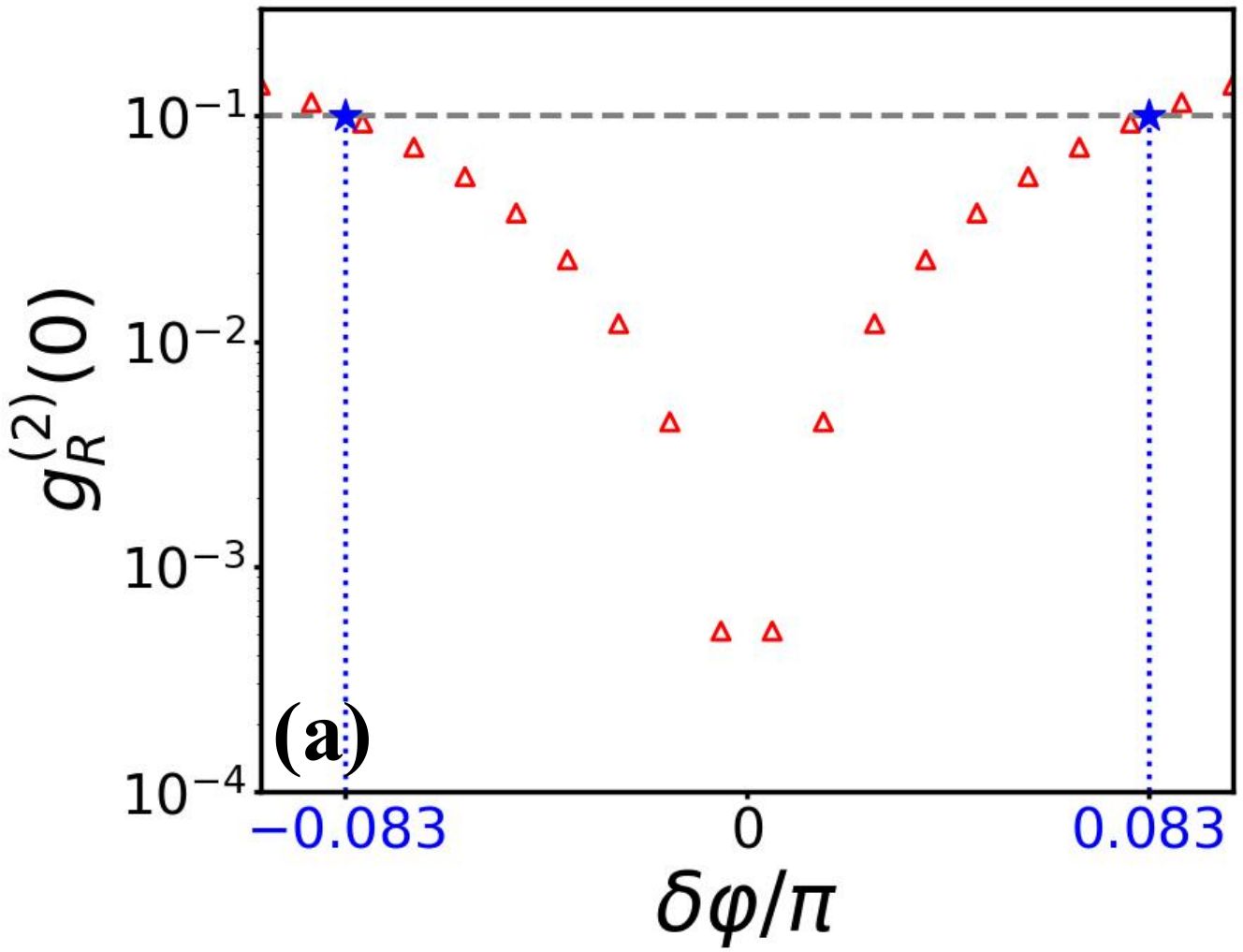}%
    \hfill
    \includegraphics[width=0.5\linewidth]{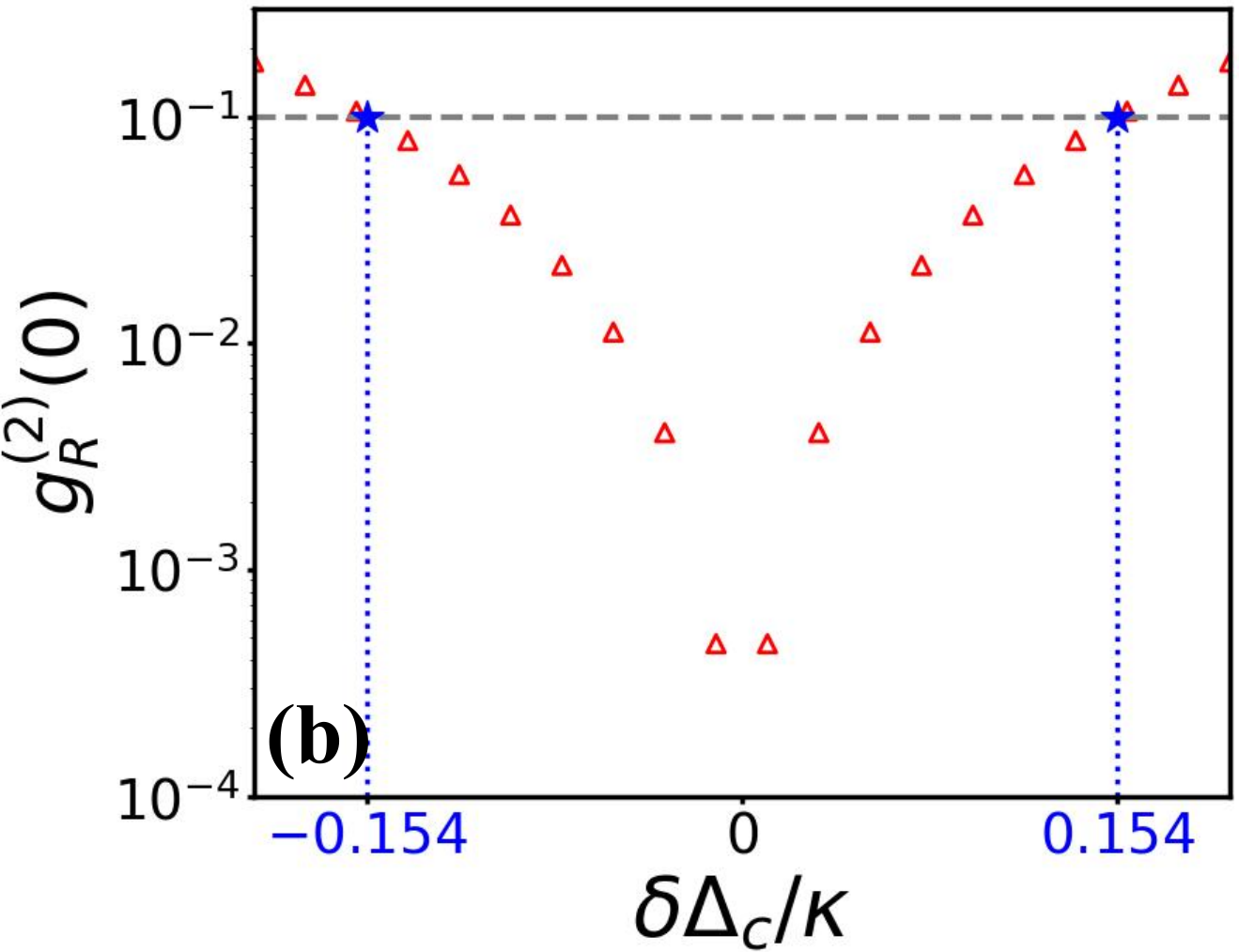}
    \includegraphics[width=0.5\linewidth]{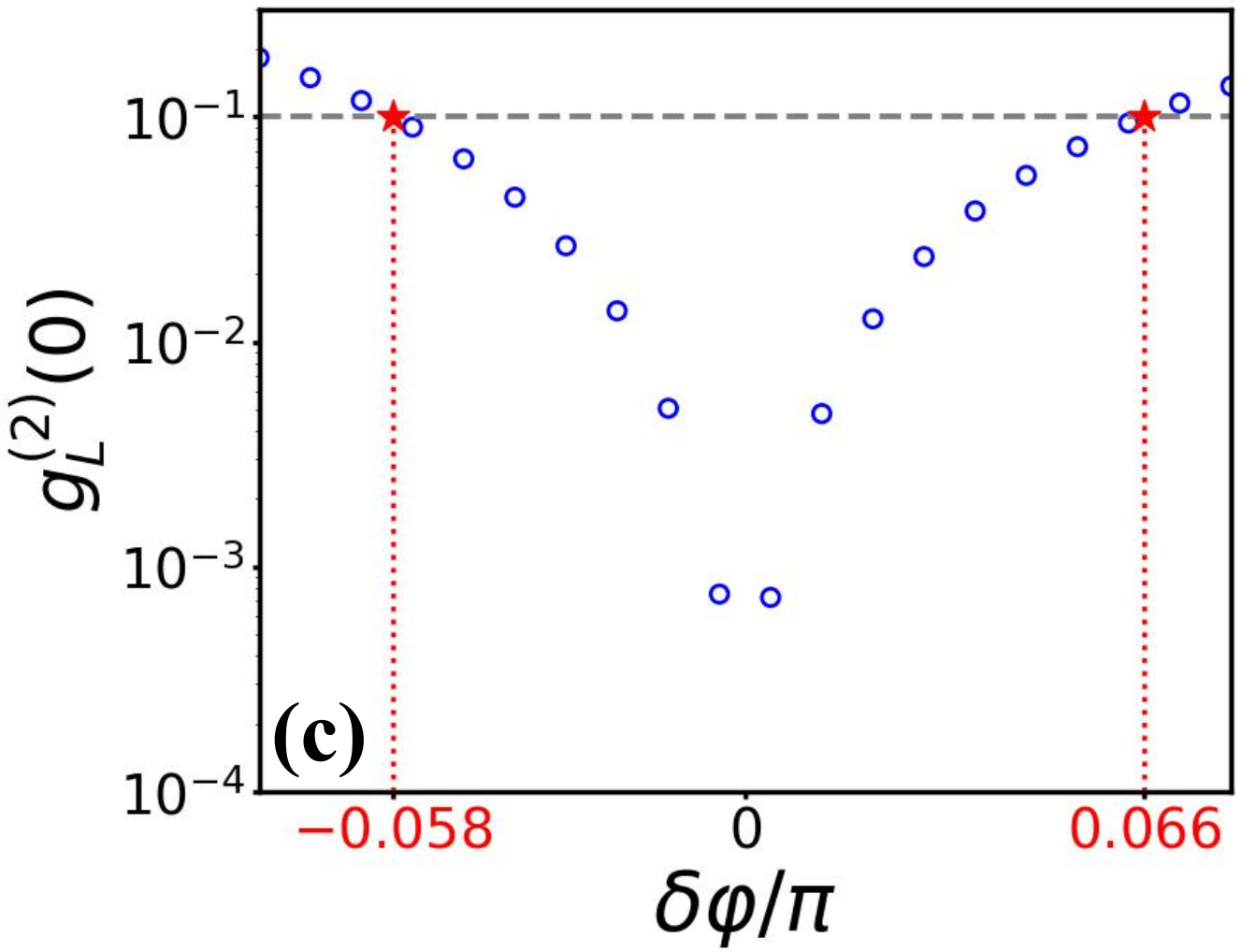}%
    \hfill
    \includegraphics[width=0.5\linewidth]{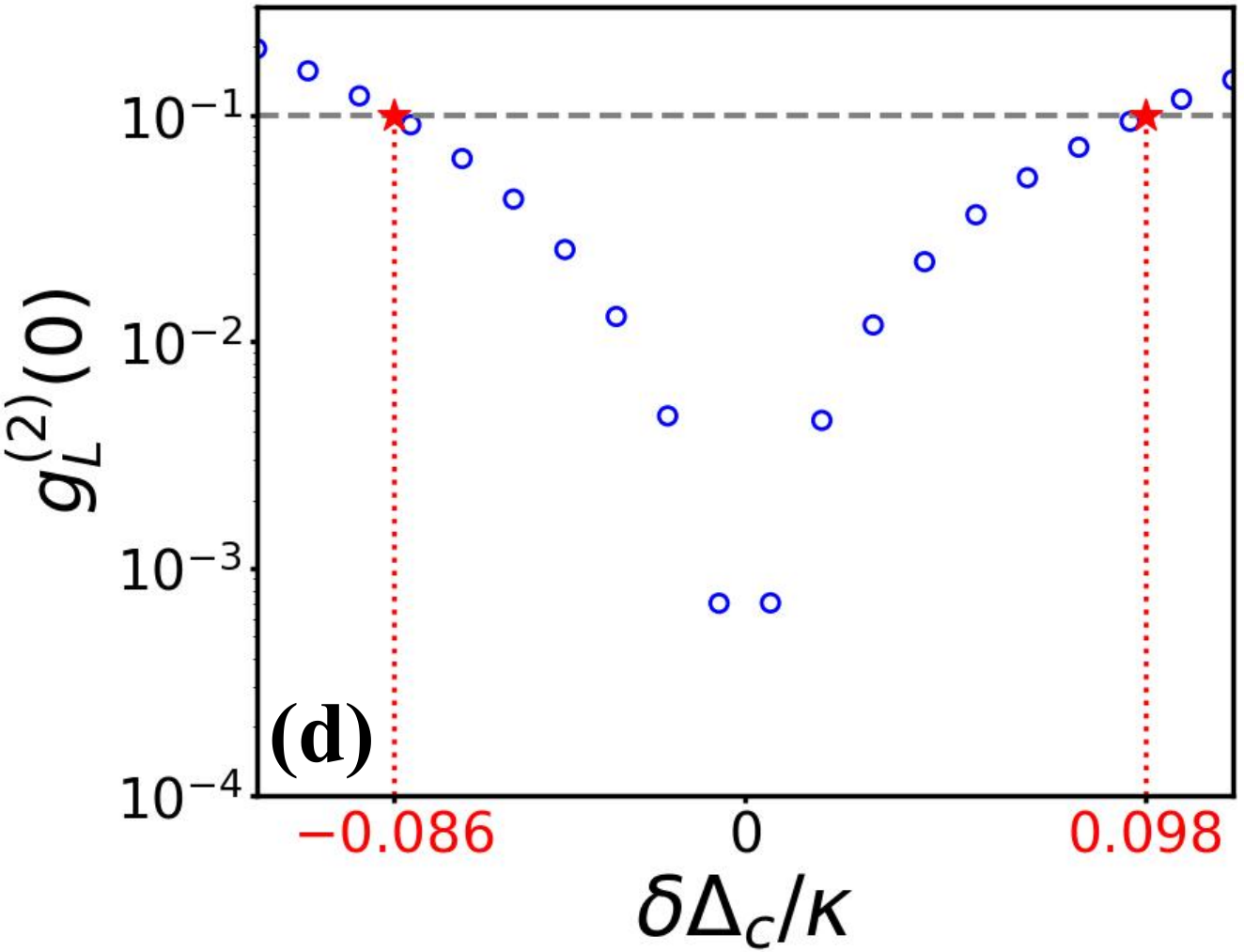}
    \caption{Numerical analysis of photon blockade robustness against parameter deviations. (a) $g_R^{(2)}(0)$ versus phase deviation $\delta\varphi/\pi$ with $g/\kappa = 0.354$ and $\Delta_c = \Delta_a/2 = 0$. (b) $g_R^{(2)}(0)$ versus cavity detuning deviation $\delta\Delta_c/\kappa$ with $g/\kappa = 0.354$ and $\varphi = 0$. (c) $g_L^{(2)}(0)$ versus phase deviation $\delta\varphi/\pi$ with $g/\kappa = 1$, $\Delta_c/\kappa = 0.866$, and $\Delta_a = 2\Delta_c$. (d) $g_L^{(2)}(0)$ versus cavity detuning deviation $\delta\Delta_c/\kappa$ with $g/\kappa = 1$, $\Delta_c/\kappa = 0.866$, $\Delta_a = 2\Delta_c$, and $\varphi = 0$. Other common parameters are $\gamma/\kappa = 1$, $\Omega_c/\kappa = 0.01$, and $|r| = 1$.}
    \label{fig7}
\end{figure}

The results presented in Figs. \ref{fig4} and   \ref{fig5} indicate that the strong PB relies strictly on precise parameter matching, making the system sensitive to parameter fluctuations. This necessitates a detailed quantitative analysis of the robustness against potential experimental imperfections. In realistic physical implementations, experimental parameters inevitably deviate from their idealized theoretical values due to fabrication imperfections and environmental fluctuations. Therefore, evaluating the tolerance space is crucial for demonstrating the experimental feasibility of our scheme.

We first investigate the influence of the phase deviation $\delta\varphi/\pi$. As shown in Fig. \ref{fig7}(a) and Fig. \ref{fig7}(c), the system can tolerate a phase deviation within the range of $[-0.083, 0.083]$ for the resonance case and $[-0.058, 0.066]$ for the off-resonance case. Physically, the relative phase is governed by $\varphi = \beta L$, where $L$ represents the distance between the two waveguide-cavity junctions. In typical WGM microcavity setups, $L$ is theoretically constrained to be less than $20\ \mu{\rm m}$, so that we can assume the photon propagation to have no time delay. Based on the extracted phase tolerance, this fluctuation translates directly into a geometric fabrication error via $\delta L = \delta\varphi/\beta = (\delta\varphi/\pi)(\lambda_{\text{eff}}/2)$, where $\lambda_{\text{eff}} = \lambda_0 / n_{\text{eff}}$ denotes the effective wavelength \cite{46}. Considering a typical resonance energy of $\omega_c = 0.78122\ \text{eV}$ (corresponding to a vacuum wavelength $\lambda_0 \approx 1587\ \text{nm}$) and an effective refractive index $n_{\text{eff}} \approx 2.5$ for the fundamental mode \cite{46}, we obtain $\lambda_{\text{eff}} \approx 635\ \text{nm}$. Substituting the aforementioned tolerance bounds, the permissible geometric deviations $\delta L$ are calculated to be $\sim 26\ \text{nm}$ for the resonance case and $\sim 18-21\ \text{nm}$ for the off-resonance case. Given that the current positioning accuracy for a quantum dot can be up to $\sim20$ nm based on fluorescence imaging \cite{46,69}, the spatial errors can be strictly controlled well within these permissible boundaries, demonstrating that our scheme exhibits high experimental feasibility on the geometric scale.

In addition, we evaluate the system's robustness against the cavity detuning deviation $\delta\Delta_c/\kappa$. As illustrated in Fig. \ref{fig7}(b) and Fig. \ref{fig7}(d), the system can endure a detuning deviation ranging from $[-0.154, 0.154]$ for the resonance case and $[-0.086, 0.098]$ for the off-resonance case, while effectively maintaining $g^{(2)}(0) < 0.1$. This numerical finding reveals that, although the tolerance bounds are slightly narrowed in the off-resonance regime, the system still exhibits substantial robustness in frequency matching. In practical solid-state cavity QED experiments, frequency mismatches typically arise from local thermal drifts or the inhomogeneous broadening of quantum emitters \cite{70}. The obtained wide tolerance intervals indicate that our scheme can comfortably accommodate these conventional spectral fluctuations, further ensuring its stable operation in realistic environments.

    \textcolor{black}{\section{photon bunching IN THE ATOM-DRIVING CASE}}

  In this section, we investigate the photon statistics of the left and right cavities by numerically calculating the equal-time second-order correlation functions of the system.

  When a monochromatic weak driving field is applied to the atom continuously, the total Hamiltonian reads ${H_t'}=H_{\rm JC}+{H_{d}'}$. In the same rotating frame, the Hamiltonian becomes
\begin{align}
H_{\rm sys}'&=\Delta_{c}'c_{L}^{\dagger}c_L+\Delta_{c}'c_{R}^{\dagger}c_R+\Delta_a'\sigma_{+}\sigma_{-}+g(c_{L}^{\dagger2}\sigma_{-}+h.c.) \nonumber\\ 
&+g(c_{R}^{\dagger2}\sigma_{-}+h.c.)+\Omega_a(\sigma_{+}+\sigma_{-}),
		\label{Eq.21}
     \end{align}
where $\Delta_c'=\omega_c-\omega_L/2$ ($\Delta_a'=\omega_a-\omega_L$) is the detuning of the cavity-field (atomic) frequency with respect to the driving frequency. Consistent with   Sect. \ref{3h}, $\omega_c=\omega_a/2$ is set,  then we have $\Delta_c'=\Delta_a'/2$.

By substituting $H_{\rm sys}'$ for $H_{\rm sys}$ in Eq. (\ref{Eq.10}), we derive the effective Hamiltonian for this system
\begin{align}
		H_{\rm eff}'=H_{\rm sys}'-\sum_{s=L,R}^{}i\frac{\kappa }{2}c_{s}^{\dagger }c_s  -i\kappa|r|e^{i\varphi}c_{L}c_{R}^{\dagger}-i\frac{\gamma}{2}\sigma_{+}\sigma_{-}.
		\label{Eq.22}
     \end{align}

We still truncate the Hilbert space of photon numbers to $n=2$ and derive the set of equations of motion for the probability amplitudes:
\begin{subequations}
\begin{align}		
  i{{\dot C}_{00g}} &= {\Omega _a}{C_{00e}},\label{25.a}\\ 
   i{{\dot C}_{10g}} &= {{\tilde \Delta }_c'}{C_{10g}}, \label{25.b} \\ 
  i{{\dot C}_{01g}} &= - i\kappa |r|{e^{i\varphi} }{C_{10g}} + {{\tilde \Delta }_c'}{C_{01g}}, \label{25.c}\\ 
  i{{\dot C}_{11g}} &= 2{{\tilde \Delta }_c'}{C_{11g}}  - \sqrt 2i\kappa |r|{e^{i\varphi} }{C_{20g}},\label{25.d}\\ 
  i{{\dot C}_{20g}} &= 2{{\tilde \Delta }_c'}{C_{20g}}+\sqrt2gC_{00e},\label{25.e}\\ 
 i{{\dot C}_{02g}}&=-\sqrt 2i\kappa |r|{e^{i\varphi} }{C_{11g}}+2{{\tilde \Delta }_c'}C_{02g}+\sqrt{2}gC_{00e},\label{25.f}\\
  i{{\dot C}_{00e}} &=\Omega_aC_{00g}+\sqrt{2}gC_{20g}+\sqrt{2}gC_{02g}+\tilde{\Delta}_a'C_{00e}.\label{25.g}   
  \end{align}
  \end{subequations}
Here, we define $\tilde{\Delta_c'}=\Delta_c'-i\kappa/2$ and
$\tilde{\Delta_a'}=\Delta_a'-i\gamma/2$. Under the steady-state approximation, we obtain $C_{10g}=C_{01g}=0$. This result implies that the excitations of the single-photon states in both the left and right cavities are completely suppressed. To provide a physical explanation for this suppression, as can be seen from Fig. \ref{fig2}(c), due to the two-photon interaction, there exist no transition pathways that allow the system to transition from the initial state $\vert{}00g\rangle$ to the single-photon states $\vert{}01g\rangle$ or $\vert{}10g\rangle$. In contrast, the excitation of the two-photon states remains unsuppressed, because there is exactly one pathway in either the left or right cavity to reach the two-photon Fock state, which can be expressed as $\vert{}00g\rangle \xrightarrow{\Omega_a} \vert{}00e\rangle \xrightarrow{g} \vert{}20g\rangle$ and $\vert{}00g\rangle \xrightarrow{\Omega_a} \vert{}00e\rangle \xrightarrow{g} \vert{}02g\rangle$, respectively. As a result, due to that there is only a single pathway available, quantum interference cannot occur. According to the expression of $g_s^{(2)}(0)$ ($s=L,R$), its value approaches infinity, signifying the occurrence of the photon bunching effect in both cavities.

\textcolor{black}{ \section{SUMMARY AND CONCLUSION}}
 
In this work, we systematically investigate the physical mechanisms for achieving universal photon blockade in a nonreciprocal microcavity system modulated by CEPs under either monochromatic weak cavity driving or atom driving.
 Through a combination of analytical derivation and numerical simulation, we demonstrate that under the cavity-driving scheme, \textcolor{black}{ the photon statistics in the two cavities exhibit remarkable differences upon introducing the CEPs: when the driving field is resonant with the cavity frequency, the right cavity (CW mode) exhibits a pronounced PB effect, while the left cavity (CCW mode) maintains a Poissonian distribution;} Notably, under off-resonant conditions, the left cavity (CCW mode) exhibits PB while the right cavity (CW mode) exhibits photon bunching. In contrast, the atom-driving scheme limits the effective excitation of single-photon states, leading to photon bunching  in both cavities.
Our results not only confirm the feasibility of modulating high-performance non-classical light sources within the weak nonlinearity regime but also highlight the significant potential of non-Hermitian EPs for the development of integrated, nonreciprocal quantum photonic chips.

\vskip 0.5cm
	
	\begin{acknowledgements}
We thank Prof. H.-Z. Shen for his helpful discussions during the revision. This work was supported by the National Science Foundation of
China (Grant nos. 12475009, 12075001, and 62471001), Anhui Provincial University Scientific Research Major Project (Grant No. 2024AH040008), the Anhui Provincial Natural Science Foundation (Grant No. 2508085ZD001), and Anhui Province Science and Technology Innovation Project (Grant No.
202423r06050004).
    
	\end{acknowledgements}

\end{document}